\documentclass[twocolumn,aps,pre,amsmath,longbibliography,nofootinbib]{revtex4-2}
\usepackage{amsmath,amssymb}
\usepackage{graphicx}
\usepackage[colorlinks=true, bookmarks=true,
bookmarksnumbered=true, bookmarkstype=toc, linkcolor=magenta,
urlcolor=cyan, citecolor=cyan]{hyperref}

\begin{document}
\title{Anomalous distribution of magnetization\\
in an Ising spin glass with correlated disorder}
\author{Hidetoshi Nishimori}
\affiliation{Institute of Integrated Research, Institute of Science Tokyo, Nagatsuta-cho, Midori-ku, Yokohama 226-8503, Japan}
\affiliation{Graduate School of Information Sciences, Tohoku University, Sendai 980-8579,
Japan}
\affiliation{RIKEN Interdisciplinary Theoretical and Mathematical Sciences (iTHEMS),
Wako, Saitama 351-0198, Japan}
. \date{\today}
\begin{abstract} 
The effect of correlations in disorder variables is a largely unexplored topic in spin glass theory. We study this problem through a specific example of correlated disorder introduced in the Ising spin glass model. We prove that the distribution function of the magnetization along the Nishimori line in the present model is identical to the distribution function of the spin glass order parameter in the standard Edwards-Anderson model with symmetrically-distributed independent disorder. This result means that if the Edwards-Anderson model exhibits replica symmetry breaking, the magnetization distribution in the correlated model has support on a finite interval, in sharp contrast to the conventional understanding that the magnetization distribution has at most two delta peaks. This unusual behavior challenges the traditional argument against replica symmetry breaking on the Nishimori line in the Edwards-Anderson model. In addition, we show that when temperature chaos is present in the Edwards-Anderson model, the ferromagnetic phase is strictly confined to the Nishimori line in the present model. These findings are valid not only for finite-dimensional systems but also for the infinite-range model, and highlight the need for a deeper understanding of disorder correlations in spin glass systems.
\end{abstract}
\maketitle

\section{Introduction}
\label{sec:introduction}
Spin glasses are among the most challenging problems in statistical physics with applications that extend far beyond the traditional boundaries of the field \cite{Nishimori2001,Huang2022,Charbonneau2023}. Exact solutions for spin glass models beyond the mean-field theory \cite{Sherrington1975,Parisi1980} are nearly non-existent, with the exception of the exact expression for energy in a subspace of the phase diagram in arbitrary dimensions, derived using gauge symmetry \cite{Nishimori1981,Nishimori1980,Nishimori2001}.

Most theoretical and numerical studies, including these mean-field and gauge-invariant solutions, start with the assumption of a spatially uncorrelated distribution of bond disorder as first proposed by Edwards and Anderson \cite{Edwards1975}. The rationale behind this assumption is that unless the correlations in disorder are strong, they would not affect the universal features of the spin glass transition. While this idea generally holds true according to the concept of the renormalization group as confirmed in many examples \cite{Nogueira1999,Honecker2001,Jorg2006,Hasenbusch2008,ParisenToldin2010,Jorg2006,Katzgraber2006,Katzgraber2007,Hasenbusch2008a}, it is important to rigorously verify its limitations. This problem is relevant because in spin glass materials, disorder in the bond variables often arises from site disorder \cite{Mydosh1993}, which can result in spatial correlations. Also, in the field of quantum error correction, to which the theoretical framework developed in the spin glass theory using gauge symmetry \cite{Nishimori1980,Nishimori1981,Nishimori2001} has been applied \cite{Dennis2002,Wang2002,Bombin2012,
Ohzeki2012,Andrist2015,fujii2015,Andrist2016,Kubina2018,Lee2023,Florian2023,Fan2024,Chen2024,Chen2024a,Dua2024,Behrends2024,Takada2024,hauser2024information,Su2024,Zhang2024,lee2024,xiao2024,li2024,wang2024a,niwa2024}, it is an interesting and important question how correlations between errors in qubits affect the performance of error correcting codes \cite{Clemens2004,Klesse2005,Aharonov2008,Preskill2013,Ball2016,Wilen2021,Chubb2021,McEwen2021,Zou2024,Postema2024}.

However, analyzing the effects of spatial correlations in disorder is difficult, and only a limited number of papers have addressed this issue \cite{Hoyos2011,Bonzom2013,Cavaliere_2019,Munster2021,Nishimori2022}. The general conclusion is that if the correlations in disorder are weak, they do not result in qualitatively different behaviors compared to the case without correlations.

In this paper, we introduce relatively strong spatial correlations between disorder variables to see if and how such correlations change the system properties.  A subspace of the phase diagram, often referred to as the Nishimori line (NL) \cite{Honecker2001,Dennis2002,Wang2002,Hasenbusch2008,Hasenbusch2008a,ParisenToldin2010,Bombin2012,Ohzeki2012,Andrist2015,fujii2015,Andrist2016,Kubina2018,alberici2021multi,tanaka2022nishimori,garban2022continuous,Lee2023,Florian2023,okuyama2023,okuyama2023mean,itoi2023boundedness,Placke2023,terasawa2023dynamical,Munster2023,camilli2023onset,
itoi2024gauge,sirenko2024paradigm,Agrawal2023,Braunstein2023,crotti2023matrix,ghio2024sampling,genovese2023minimax,ettori2023finite,Angelini2023,
Fan2024,Chen2024,Chen2024a,Dua2024,Behrends2024,Takada2024,hauser2024information,Su2024,Zhang2024,lee2024,xiao2024,li2024,wang2024a,niwa2024,Zdeborova2016,Zhu2023,chen2023,lee2022}, plays a crucial role in the analysis.  We prove that the distribution function of the magnetization on the NL is equal to the distribution function of the spin glass order parameter in the Edwards-Anderson model with symmetrically-distributed independent disorder. It immediately follows that if the latter distribution for spin glass ordering has support on a finite interval reflecting replica symmetry breaking, then the former distribution for the magnetization on the NL with correlated disorder has also the same support of a finite interval. This is anomalous because it has traditionally been believed that the distribution of the magnetization of the Ising spin glass has only two delta peaks in the ferromagnetic phase, based on which the argument for the absence of replica symmetry breaking on the NL for the Edwards-Anderson model has been developed \cite{Nishimori2001b,Nishimori2001}. Also proved is that the distribution function of the magnetization at an arbitrary point in the phase diagram except for the NL is identical to the distribution function of the overlap of two replicas at different temperatures in the symmetrically-distributed Edwards-Anderson model. One of the important consequences of this identity is that the ferromagnetic phase is confined to the NL in the phase diagram if temperature chaos exists in the Edwards-Anderson model. These unconventional properties are the natural consequences of the specific type of correlations in disorder.

The next section is the main part of the paper, where the problem is defined and analyzed by the method of gauge transformation.  Summary and discussions are given in the last section.

\section{Model and its analyses}
\label{sec:main}

\subsection{Problem definition}
We study the Ising spin glass with the Hamiltonian
\begin{align}
    H=-J \sum_{\langle ij\rangle}\tau_{ij}S_i S_j\quad (S_i=\pm 1).
\end{align}
The summation runs over all interacting spin pairs on an arbitrary lattice with an arbitrary range of interactions. The coupling constant $J$ will be chosen to be unity $J=1$ without losing generality. Disorder is represented by the quenched random variables $\tau_{ij}(=\pm 1)$ with the distribution function
\begin{align}
    P(\tau)=\frac{1}{A}\, \frac{e^{\beta_p \sum_{\langle ij\rangle} \tau_{ij}}}{Z_{\tau}(\beta_p)},
    \label{eq:P_correlated}
\end{align}
where $\beta_p$ is a parameter to control the properties of the distribution, and $Z_{\tau}(\beta_p)$ is the partition function for a fixed configuration of $\tau=\{\tau_{ij}\}$,
\begin{align}
    Z_{\tau}(\beta_p)=\sum_S e^{\beta_p \sum_{\langle ij\rangle} \tau_{ij}S_i S_j}.
\end{align}
The normalization factor $A$ is evaluated as
\begin{align}
    A&=\sum_{\tau}\frac{e^{\beta_p \sum_{\langle ij\rangle} \tau_{ij}}}{Z_{\tau}(\beta_p)}=\frac{1}{2^N}\,\sum_{\tau}\frac{\sum_{\sigma}e^{\beta_p \sum_{\langle ij\rangle} \tau_{ij}\sigma_i\sigma_j}}{Z_{\tau}(\beta_p)}\nonumber\\
    &=\frac{2^{N_{\rm B}}}{2^N}.
    \label{eq:Ap}
\end{align}
In the second equality, we applied the gauge transformation $\tau_{ij}\to \tau_{ij}\sigma_i\sigma_j$ $(\sigma_i, \sigma_j=\pm 1)$, summed the resulting numerator over all values of $\sigma$, and divided the result by $2^N$. $N_{\rm B}$ is the number of bonds (interacting pairs) and $N$ is the number of sites.
Equation (\ref{eq:P_correlated}) is not decomposed into the product of independent distribution for each $\tau_{ij}$ unlike in the Edwards-Anderson model with
\begin{align}
    P_{\rm EA}(\tau)=\frac{e^{\beta_p\sum_{\langle ij\rangle}\tau_{ij}}}{(2\cosh \beta_p)^{N_{\rm B}}}
\end{align}
and therefore $P(\tau)$ represents correlated disorder. Note that $P_{\rm EA}(\tau)$ represents the $\pm J$ model with the probability $p$ for ferromagnetic bonds and $1-p$ for antiferromagnetic bonds by the relation $e^{-2\beta_p}=(1-p)/p$.

The probability distribution $P(\tau)$ is likely to have an enhanced weight for bond configurations with strong frustration as compared to the Edwards-Anderson model $P_{\rm EA}(\tau)$ without $Z_{\tau}(\beta_p)$ in the denominator. The reason is that the free energy
\begin{align}
    \beta_p F_{\tau}(\beta_p)=-\ln Z_{\tau}(\beta_p)
    \label{eq:Free-energy}
\end{align}
is expected to have a higher value (thus a smaller value of $Z_{\tau}(\beta_p)$) when frustration is stronger because a system with stronger frustration may be less stable than a system with weaker frustration. This means that  the denominator $Z_{\tau}(\beta_p)$ in Eq.~(\ref{eq:P_correlated}) enhances the probability for the case with stronger frustration.

The probability distribution $P(\tau)$ reduces to that of the Edwards-Anderson model with symmetric (unbiased) distribution in the limit $\beta_p\to 0$,
\begin{align}
    \lim_{\beta_p\to 0}P(\tau)=\lim_{\beta_p\to 0}P_{\rm EA}(\tau)=\frac{1}{2^{N_{\rm B}}}.
\end{align}
In the opposite limit $\beta_p\to\infty$, $P(\tau)$ allows for a number of non-trivial configurations of the disorder variables and does not reduce to the pure ferromagnetic model, thus differing from the Edwards-Anderson model,
\begin{align}
    \lim_{\beta_p\to \infty}P(\tau)\ne\lim_{\beta_p\to \infty}P_{\rm EA}(\tau).
\end{align}
A simple example of the disorder configuration for $P(\tau)$ in this limit is illustrated in Fig.~\ref{fig1}. Additional information on the properties of $P(\tau)$ is provided in the Appendix.

Note that we keep the system size $N$ finite when we take various limits such as $\beta_p\to \infty$. The thermodynamic limit $N\to\infty$ is taken after all other manipulations have been carried out.
\begin{figure}[t]
\centering
\includegraphics[width=50mm]{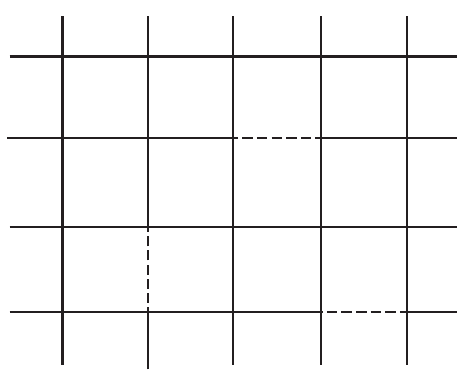}
\caption{One of the bond configurations that survives in the limit $\beta_p\to\infty$. Full lines represent ferromagnetic bonds $\tau_{ij}=1$, and dotted lines are for antiferromagnetic bonds $\tau_{ij}=-1$. The ground state for this bond configuration has all spins aligned up (or all down).
}
\label{fig1}
\end{figure}

The thermal average with inverse temperature $\beta$ for a given configuration of $\tau_{ij}$ will be denoted as $\langle \cdots \rangle_{\beta}$, and the configurational average over the bond variables by the probability distribution $P(\tau)$ is written as $[\cdots]_{\beta_p}$. The NL is defined by $\beta_p=\beta$ as in the Edwards-Anderson model.

The analyses below apply also to the case of continuous couplings with Gaussian weight.  One simply replaces $\tau_{ij}$ by $J_{ij}$, and the summation over $\tau$ variables is replaced by the integral over $J_{ij}$,
\begin{align}
    \frac{1}{2^{N_{\rm B}}}\sum_{\tau}(\cdots) \longrightarrow \int_{-\infty}^{\infty} (\cdots)\prod_{\langle ij\rangle} \frac{e^{-\frac{{J_{ij}}^2}{2}}}{\sqrt{2\pi}}\, dJ_{ij}. \label{eq:gaussian}
\end{align}

\subsection{Gauge invariant quantities}
A first simple observation is that the configurational average of a gauge-invariant quantity $Q_{\tau}(\beta)$ such as the energy and the spin glass order parameter does not depend on $\beta_p$, 
\begin{align}
    [Q_{\tau}(\beta)]_{\beta_p}&=\frac{1}{A}\sum_{\tau}\frac{e^{\beta_p\sum \tau_{ij}}}{Z_{\tau}(\beta_p)}\, Q_{\tau}(\beta)\nonumber\\
    &=\frac{1}{2^N A}\sum_{\tau}\frac{\sum_{\sigma}e^{\beta_p\sum \tau_{ij}\sigma_i\sigma_j}}{Z_{\tau}(\beta_p)}\, Q_{\tau}(\beta)\nonumber\\
    &=\frac{1}{2^{N_{\rm B}}}\sum_{\tau}Q_{\tau}(\beta)=[Q_{\tau}(\beta)]_0.
\end{align}
For example, the spin glass order parameter satisfies $[\langle S_i\rangle_{\beta}^2]_{\beta_p}=[\langle S_i\rangle_{\beta}^2]_0$, where $\langle S_i\rangle_{\beta}$ is computed under fixed boundary conditions to avoid trivial vanishing. The right-hand side $[\langle S_i\rangle_{\beta}^2]_0$ is the spin glass order parameter of the Edwards-Anderson model with symmetric distribution. Therefore, the critical line marking the onset of finite spin glass ordering (and ferromagnetic ordering) is a straight horizontal line spanning all values of $\beta_p$ in the phase diagram drawn in terms of $\beta_p$ and $\beta$. See Figs. \ref{fig:phase_diagram1} and \ref{fig:phase_diagram2} below.

\subsection{Distribution functions}
\subsubsection{Definition and the main identity}
The distribution functions of the magnetization $P_1(x|\beta,\beta_p)$ and the replica overlap $P_2(x|\beta_1,\beta_2)$ are defined as follows:
\begin{widetext}
\begin{align}
    P_1(x|\beta,\beta_p)
    &=\frac{1}{A}\sum_{\tau}\frac{e^{\beta_p \sum \tau_{ij}}}{Z_{\tau}(\beta_p)}
   \,
    \frac{\sum_S \delta\big(x-\frac{1}{N}\sum_iS_i\big)e^{\beta \sum \tau_{ij}S_iS_j}}{\sum_S e^{\beta \sum \tau_{ij}S_iS_j}}.\label{eq:P1}\\
    P_2(x|\beta_1,\beta_2)
    &=\frac{1}{A}\sum_{\tau}\frac{e^{\beta_p \sum \tau_{ij}}}{Z_{\tau}(\beta_p)}
    \,\frac{\sum_{S^{(1,2)}} \delta\big(x-\frac{1}{N}\sum_iS_i^{(1)}S_i^{(2)}\big)e^{\beta_1 \sum \tau_{ij}S_i^{(1)}S_j^{(1)}}e^{\beta_2 \sum \tau_{ij}S_i^{(2)}S_j^{(2)}}}{\sum_{S^{(1,2)}} e^{\beta_1 \sum \tau_{ij}S_i^{(1)}S_j^{(1)}}e^{\beta_2 \sum \tau_{ij}S_i^{(2)}S_j^{(2)}}}\nonumber\\
    &=\frac{1}{2^{N_{\rm B}}}\sum_{\tau}
    \frac{\sum_{S^{(1,2)}} \delta\big(x-\frac{1}{N}\sum_iS_i^{(1)}S_i^{(2)}\big)e^{\beta_1 \sum \tau_{ij}S_i^{(1)}S_j^{(1)}}e^{\beta_2 \sum \tau_{ij}S_i^{(2)}S_j^{(2)}}}{\sum_{S^{(1,2)}} e^{\beta_1 \sum \tau_{ij}S_i^{(1)}S_j^{(1)}}e^{\beta_2 \sum \tau_{ij}S_i^{(2)}S_j^{(2)}}}.
    \label{eq:P2}
\end{align}
\end{widetext}
In deriving the last expression, we have used the fact that a gauge invariant quantity does not depend on $\beta_p$, and therefore $P_2(x|\beta_1,\beta_2)$ does not carry the argument $\beta_p$.
The notation for $P_1(x|\beta,\beta_p)$ and $P_2(x|\beta_1,\beta_2)$ with a vertical bar after $x$ in the argument has been adopted to indicate that those are functions of $x$ under the given values of the hyperparameters $\beta$ and $\beta_p$ or $\beta_1$ and $\beta_2$.

Applying the gauge transformations $\tau_{ij}\to\tau_{ij}\sigma_i\sigma_j$ and $S_i\to S_i\sigma_i$ to $P_1(x|\beta,\beta_p)$ and summing the result over the $\sigma$ variables gives the following relationship,
\begin{align}
    P_1(x|\beta,\beta_p)=P_2(x|\beta,\beta_p).
    \label{eq:p1=p2}
\end{align}
This equation shows that the distribution function of the magnetization is equal to the distribution function of the replica overlap at any point in the phase diagram $(\beta,\beta_p)$. It is useful to note that $P_2(x|\beta,\beta_p)$ represents the overlap of two replicas with inverse temperatures $\beta$ and $\beta_p$ in the Edwards-Anderson model with symmetric distribution as seen in Eq.~(\ref{eq:P2}). Equation (\ref{eq:p1=p2}) is the main identity of this paper, from which several important conclusions will be drawn.

\subsubsection{Vertical phase boundary}
\label{subsub:vertical}
\begin{figure}[htbp]
 \centering
 \includegraphics[width=7cm]{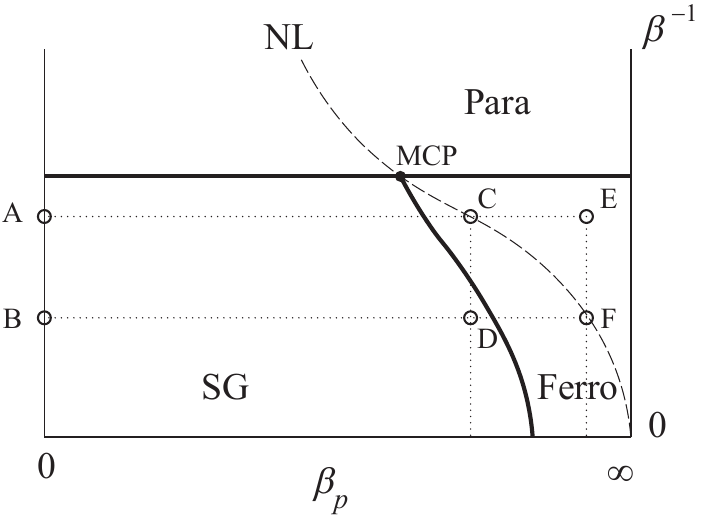}
 \caption{Phase diagram with the spin glass phase under the ferromagnetic phase. NL is the Nishimori line $\beta=\beta_p$, and MCP is the multicritical point where three phases merge. The left edge $\beta_p=0$ corresponds to the Edwards-Anderson model with uncorrelated symmetrically-distributed disorder. We assume the existence of the spin glass phase in the latter model. Otherwise, the entire phase diagram is occupied by the paramagnetic phase.}
 \label{fig:phase_diagram1}
\end{figure}

Equation (\ref{eq:p1=p2}) gives restrictions on the shape of the phase diagram. Let us suppose that the phase diagram has a spin glass phase below the ferromagnetic phase, reentrance, as illustrated in Fig.~\ref{fig:phase_diagram1}. 
Equation (\ref{eq:p1=p2}) shows, for the distribution function of the magnetization at point D in the phase diagram of Fig.~\ref{fig:phase_diagram1},
\begin{align}
 P_1(x|\beta^{\rm (D)},{\beta_p}^{({\rm D})})&=P_1(x|\beta^{\rm (D)},{\beta_p}^{\rm (C)})\nonumber\\
 &=P_2(x|\beta^{\rm (B)},{\beta}^{({\rm A})}),\label{eq:id1}
\end{align}
where we have used  $\beta_p^{\rm (D)}=\beta_p^{\rm (C)}$, $\beta^{\rm (D)}=\beta^{\rm (B)}$ and $\beta_p^{\rm (C)}=\beta^{\rm (C)}=\beta^{\rm (A)}$. Notice that C is on the NL and thus $\beta_p^{\rm (C)}=\beta^{\rm (C)}$.
Similarly, the distribution function of the magnetization at point E satisfies
\begin{align}
 P_1(x|\beta^{\rm (E)},{\beta_p}^{({\rm E})})&=P_1(x|\beta^{\rm (E)},{\beta_p}^{\rm (F)})\nonumber\\
 &=P_2(x|\beta^{\rm (A)},{\beta}^{({\rm B})}),\label{eq:id2}
\end{align}
using $\beta_p^{\rm (E)}=\beta_p^{\rm (F)}$, $\beta^{\rm (E)}=\beta^{\rm (A)}$ and $\beta_p^{\rm (F)}=\beta^{\rm (F)}=\beta^{\rm (B)}$.
Since $P_2(x|\beta^{\rm (B)},{\beta}^{({\rm A})})=P_2(x|\beta^{\rm (A)},{\beta}^{({\rm B})})$, we find from Eqs. (\ref{eq:id1}) and (\ref{eq:id2}) that the distribution functions of the magnetization at points D and E are equal to each other,
\begin{align}
    P_1(x|\beta^{\rm (D)},{\beta_p}^{({\rm D})})=P_1(x|\beta^{\rm (E)},{\beta_p}^{({\rm E})}). \label{eq:p1=p1}
\end{align}
It may be useful to note that this equation can be written in a more general form as
\begin{align}
    P_1(x|\beta,\beta_p)=P_1(x|\beta_{p},\beta).
\end{align}
The structure of the phase diagram of Fig.~\ref{fig:phase_diagram1} is not compatible with Eq.~(\ref{eq:p1=p1}) because D is in the spin glass phase with vanishing magnetization and E is in the ferromagnetic phase
\begin{align}
    &P_1(x|\beta^{\rm (D)},{\beta_p}^{({\rm D})})=\delta(x)\nonumber\\
    &P_1(x|\beta^{\rm (E)},{\beta_p}^{({\rm E})})\ne \delta(x).
\end{align}
This means that those two points do not share the same distribution function of the magnetization, a contradiction to Eq.~(\ref{eq:p1=p1}). Therefore, unlike Fig.~\ref{fig:phase_diagram1}, there should be no reentrance.

\begin{figure}[t]
 \includegraphics[width=7cm]{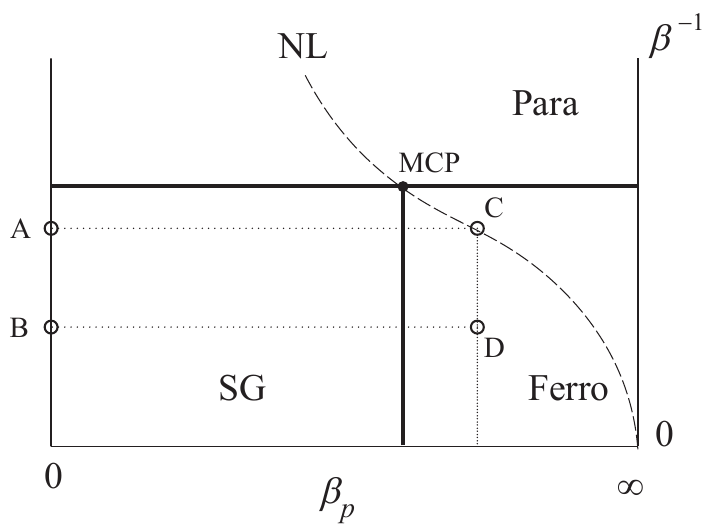}
 \caption{Phase diagram with a straight vertical boundary between the spin glass and ferromagnetic phases. NL is the Nishimori line $\beta=\beta_p$, on which the multicritical point (MCP) lies. The left edge $\beta_p=0$ corresponds to the Edwards-Anderson model with uncorrelated symmetrically-distributed disorder}
 \label{fig:phase_diagram2}
\end{figure}

The same argument applies to a possible phase diagram with the ferromagnetic phase lying below the spin glass phase.  We conclude that the boundary between the spin glass and ferromagnetic phases is a vertical line as depicted in Fig.~\ref{fig:phase_diagram2}. 

We note that gauge-invariant quantities including the spin glass order parameter and the energy have no singularities even at the vertical phase boundary since their values do not depend on $\beta_p$. This is  anomalous since, generally, physical quantities are singular at a critical point.

\subsubsection{Multicritical point on the Nishimori line}

\begin{figure}[t]
 \includegraphics[width=7cm]{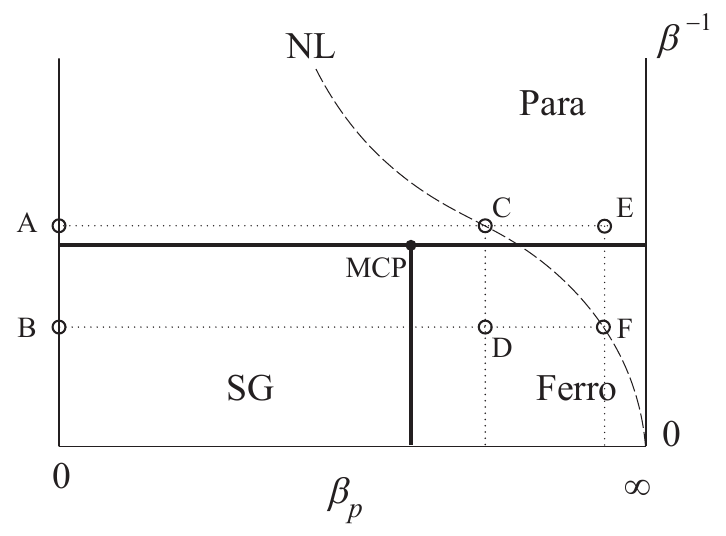}
 \caption{Phase diagram when the multicritical point (MCP) lies below  the NL $\beta=\beta_p$.  This structure is not allowed.}
 \label{fig:phase_diagram2b}
\end{figure}

It is possible to show that the multicritical point, where paramagnetic, ferromagnetic and spin glass phases merge, lies on the NL as illustrated in Fig.~\ref{fig:phase_diagram2}. Let us first point out that the spin glass order parameter $q(\beta,\beta_p)=[\langle S_i\rangle_{\beta}^2]_{\beta_p}$ is equal to the ferromagnetic order parameter $m(\beta,\beta_p)=[\langle S_i\rangle_{\beta}]_{\beta_p}$ on the NL, $q(\beta,\beta)=m(\beta,\beta)$. This relation is derived by taking the first moment (average) of both sides of Eq.~(\ref{eq:p1=p2}) and setting $\beta_p=\beta$. Consequently, the NL does not enter the spin glass phase ($q>0, m=0)$, i.e., the multicritical point is either on or below the NL. Now, assume that the multicritical point lies below the NL as in Fig.~\ref{fig:phase_diagram2b}. The following relation with the same form as Eq.~(\ref{eq:p1=p1}) can be derived,
\begin{align}
    P_1(x|\beta^{\rm (D)},{\beta_p}^{({\rm D})})=P_1(x|\beta^{\rm (E)},{\beta_p}^{({\rm E})}). \label{eq:p1=p1_2}
\end{align}
This contradicts Fig.~\ref{fig:phase_diagram2b} since D is in the ferromagnetic phase and E is in the paramagnetic phase. It is therefore concluded that the multicritical point lies on the NL.

\subsubsection{Support on a finite interval for magnetization}
\begin{figure}[t]
\centering
\includegraphics[width=65mm]{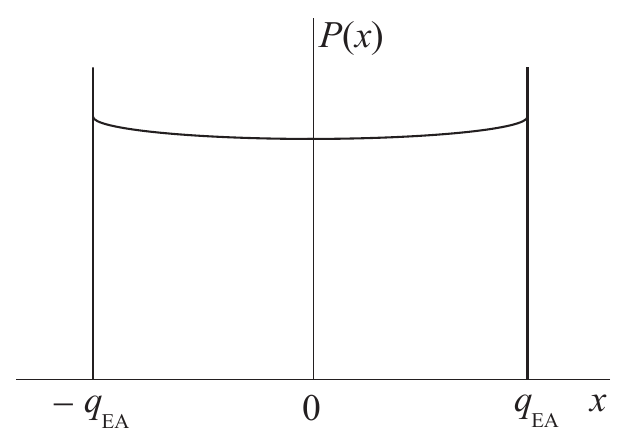}
\caption{Distribution function of the spin glass order parameter under the existence of replica symmetry breaking has support on a finite interval. $q_{\rm EA}$ is the Edwards-Anderson order parameter $[\langle S_i\rangle_{\beta}^2]_0$.}
\label{fig:q-distribution}
\end{figure}

We next discuss the distribution functions on the NL $\beta=\beta_p$. Equation (\ref{eq:p1=p2}) leads to the following identity for the distribution function of the magnetization at point C on the NL in Fig.~\ref{fig:phase_diagram2},
\begin{align}
    P_1(x|\beta^{\rm (C)},{\beta}^{({\rm C})})= P_2(x|\beta^{\rm (C)},{\beta}^{({\rm C})})
    =P_2(x|\beta^{({\rm A})},\beta^{({\rm A})}).
    \label{eq:finite_distribution_m}
\end{align}
The second equality comes from the fact that the gauge invariant $P_2(x|\beta_{1},{\beta}_{2})$ does not depend on the horizontal position in the phase diagram. The right-most expression of the above equation is the distribution function of the spin glass order parameter for the Edwards-Anderson model.  If the latter has replica symmetry breaking, the distribution function $P_2(x|\beta^{({\rm A})},\beta^{({\rm A})})$ has support on a finite interval as illustrated in Fig.~\ref{fig:q-distribution}.
Then, Eq.~(\ref{eq:finite_distribution_m}) implies that the distribution function of the magnetization at point C has exactly the same form as in Fig.~\ref{fig:q-distribution} having support on a finite interval.

This result is anomalous because the distribution function of the magnetization has been believed to have a simple form of two delta peaks at $\pm m=\pm [\langle S_i\rangle_{\beta}]_{\beta_p}$ in the ferromagnetic phase. Equation~(\ref{eq:finite_distribution_m}) challenges this traditional understanding if the right-hand side has support on a finite interval. Since the reasoning for the absence of replica symmetry breaking on the NL in the Edwards-Anderson model relies on the above-mentioned traditional viewpoint for the distribution of the magnetization \cite{Nishimori2001b,Nishimori2001}, it is crucial to carefully examine the implications of the present result.

\subsubsection{Temperature chaos}
\begin{figure}[t]
\centering
\includegraphics[width=65mm]{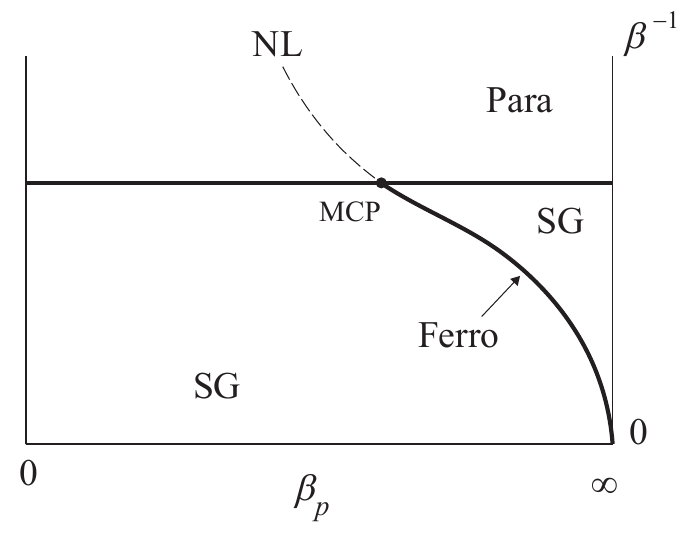}
\caption{Phase diagram with the ferromagnetic phase only on the NL is realized when the Edwards-Anderson model ($\beta_p=0$) has temperature chaos. MCP is the multicritical point.
}
\label{fig:phase_diagram3}
\end{figure}

We next consider the replica overlap $P_2(x|\beta^{\rm (A)},\beta^{\rm (B)})$ at points A and B in Fig.~\ref{fig:phase_diagram2}. 
This is a measure of temperature chaos \cite{Parisi2010}. Temperature chaos is a peculiar property of the spin glass state, where any small amount of temperature change leads to a completely different spin configuration \cite{Bray1987,Banavar1987,Fisher1986,Fisher1988, Kondor1989,Ney-Nifle_1997,Ney-Nifle1998,Parisi2010,Mathieu2001,Bouchaud2001,Aspelmeier2002,Rizzo2003,Houdayer2004,Katzgraber2007b,Fernandez_2013,Wang2015,Billoire2018,Baity-Jesi2021}. More explicitly, temperature chaos is defined to exist in the Edwards-Anderson model if the replica-overlap distribution at different temperatures is a delta function \cite{Parisi2010},
\begin{align}
    P_2(x|\beta^{\rm (A)},\beta^{\rm (B)})=\delta(x)\quad (\beta^{\rm (A)}\ne\beta^{\rm (B)}).\label{eq:tc1}
\end{align}
On the other hand, we find
\begin{align}
     &P_2(x|\beta^{\rm (B)},\beta^{\rm (A)})=P_2(x|\beta^{\rm (D)},\beta_p^{\rm (C)})\nonumber\\
     =&P_2(x|\beta^{\rm (D)},\beta_p^{\rm (D)})=P_1(x|\beta^{\rm (D)},\beta_p^{\rm (D)}),\label{eq:tc2}
\end{align}
where we have used Eq.~(\ref{eq:p1=p2}) in the last equality.
It follows from Eqs.~(\ref{eq:tc1}) and (\ref{eq:tc2}) that
\begin{align}
    P_1(x|\beta^{\rm (D)},\beta_p^{\rm (D)})=\delta(x).
\end{align}
This is in conflict with the assumption that point D is in the ferromagnetic phase.  Notice that the condition $\beta^{\rm (A)}\ne \beta^{\rm (B)}$ implies that point D is not on the NL (i.e., ${\rm C}\ne {\rm D}$). We conclude that if temperature chaos exists in the Edwards-Anderson model, there is no ferromagnetic phase except exactly on the NL in the present model. On the NL, the ferromagnetic order parameter is equal to the spin glass order parameter $m(\beta,\beta)=q(\beta,\beta)$ and thus ferromagnetic ordering exists if spin glass ordering does. See Fig.~\ref{fig:phase_diagram3}. If temperature chaos does not exist in the Edwards-Anderson model, the phase diagram of Fig.~\ref{fig:phase_diagram2} is allowed.

\subsection{Identity and inequality}
In addition to the identities on the distribution functions, we can derive the following relationships between correlation functions,
\begin{align}
 \left[\Big(\big\langle \prod_{i\in D}S_i\big\rangle_{\beta}\Big)^{2n-1}\right]_{\beta_p}&= \left[\Big(\big\langle \prod_{i\in D}S_i\big\rangle_{\beta}\Big)^{2n-1}\big\langle \prod_{i\in D}S_i \big\rangle_{\beta_p}\right]_{\beta_p} \label{eq:identity1}\\
 \left[\frac{\displaystyle\langle \prod_{i\in D}S_i\rangle_{\beta}}{\displaystyle\Big|\langle \prod_{i\in D}S_i\rangle_{\beta}\Big|}\right]_{\beta_p}&\le
 \left[\frac{\displaystyle\langle \prod_{i\in D}S_i\rangle_{\beta_p}}{\displaystyle\Big|\langle \prod_{i\in D}S_i\rangle_{\beta_p}\Big|}\right]_{\beta_p},\label{eq:identity2}
\end{align}
where $n\in \mathbb{Z}$, and $D$ is an arbitrary subset of all lattice sites. When the number of sites in $D$ is odd, fixed boundary conditions are imposed to avoid trivial vanishing of the thermal average. These are generalizations of the corresponding equations for the Edwards-Anderson model \cite{Nishimori1981,Nishimori1993,Nishimori2001}.

Equation (\ref{eq:identity1}) can be easily proved using a gauge transformation, similar to the approach used in the Edwards-Anderson model \cite{Nishimori1981}. Equation (\ref{eq:identity2}) can also be proved as in the Edwards-Anderson model \cite{Nishimori1993}, but it may be useful to recapitulate the intermediate steps, as it is a bit more involved than in the case of Eq.~(\ref{eq:identity1}).

Let us take the example of a single site in $D$ for simplicity, $D=\{i\}$. The left-hand side of Eq.~(\ref{eq:identity2}) follows:
\begin{align}
\left[\frac{\langle S_i\rangle_{\beta}}{|\langle S_i\rangle_{\beta}|}\right]_{\beta_p}&
=\frac{1}{A}\sum_{\tau} \frac{e^{\beta_p\sum \tau_{ij}}}{Z_{\tau}(\beta_p)}\frac{\langle S_i\rangle_{\beta}}{|\langle S_i\rangle_{\beta}|}
\nonumber\\
&=\frac{1}{2^NA}\sum_{\tau}\langle \sigma_i\rangle_{\beta_p}\frac{\langle S_i\rangle_{\beta}}{|\langle S_i\rangle_{\beta}|}\nonumber\\
& \le \big[|\langle \sigma_i\rangle_{\beta_p}|\big]_{0},
\label{eq:ineq1}
\end{align}
where we applied a gauge transformation in going from the second to the third expression. On the other hand,
the right-hand side of Eq.~(\ref{eq:identity2}) is, using the second line of the above equation,
\begin{align}
    \left[\frac{\langle S_i\rangle_{\beta_p}}{|\langle S_i\rangle_{\beta_p}|}\right]_{\beta_p}&=\frac{1}{2^{N_{\rm B}}}\sum_{\tau}\frac{\langle S_i\rangle_{\beta_p}^2}{|\langle S_i\rangle_{\beta_p}|}= \big[|\langle S_i\rangle_{\beta_p}|\big]_{0}.
    \label{eq:ineq2}
\end{align}
The right-hand sides of Eqs.~(\ref{eq:ineq1}) and (\ref{eq:ineq2}) coincide and therefore,
\begin{align}
    \left[\frac{\langle S_i\rangle_{\beta}}{|\langle S_i\rangle_{\beta}|}\right]_{\beta_p} \le \left[\frac{\langle S_i\rangle_{\beta_p}}{|\langle S_i\rangle_{\beta_p}|}\right]_{\beta_p}.
    \label{eq:identity2a}
\end{align}
The above derivation trivially generalizes to prove Eq.~(\ref{eq:identity2}).

Let us discuss the physical significance of these equations. Equation (\ref{eq:identity1}) for $n=1$ and $D=\{i\}$ is
\begin{align}
    [\langle S_i\rangle_{\beta}]_{\beta_p}=[\langle S_i\rangle_{\beta}\langle S_i\rangle_{\beta_p}]_{\beta_p}.
\end{align}
When $\beta_p=\beta$, this reduces to the equivalence of the ferromagnetic order parameter and the spin glass order parameter on the NL $m(\beta,\beta)=q(\beta,\beta)$, which was already derived in Sec.~\ref{subsub:vertical}. For $n=0$ and $D=\{i\}$ and $\beta=\beta_p$, we find
\begin{align}
    \left[\frac{1}{\langle S_i\rangle_{\beta}}\right]_{\beta}=1.
\end{align}
This is the same identity as in the case of the Edwards-Anderson model \cite{Nishimori2001,Nishimori1981}.

Equation (\ref{eq:identity2a}), or its representation as
\begin{align}
 \big[{\rm sgn}\big(\langle S_i\rangle_{\beta}\big)\big]_{\beta_p}\le  \big[{\rm sgn}\big(\langle S_i\rangle_{\beta_p}\big)\big]_{\beta_p},
\end{align}
where ${\rm sgn}(x)$ is $+1$ for $x>0$ and $-1$ for $x<0$,
implies that the average direction of the spins takes the maximum value on the NL as a function of $\beta$ for a given value of $\beta_p$. In other words, the spin alignment along a given (e.g. positive) direction reaches its maximum on the NL as we change the temperature.

\section{Summary and discussion}
We have introduced and analyzed an Ising spin glass with correlated disorder of a specific type.  The result shows that the distribution function of the magnetization at an arbitrary point $(\beta,\beta_p)$ in the phase diagram is equal to the distribution function of replica overlap of the Edward-Anderson model with symmetric disorder at two inverse temperatures $\beta$ and $\beta_p$. It follows that the phase boundary between the spin glass and ferromagnetic phases is a vertical line unless temperature chaos exists in the Edwards-Anderson model. If temperature chaos is present, the ferromagnetic phase is confined strictly to the NL of the present model. Also, if replica symmetry breaking exists in the Edwards-Anderson model, the distribution function of the magnetization on the NL has support on a finite interval. This latter possibility is in conflict with the traditional understanding that the distribution function of the magnetization has at most two delta peaks, based on which the absence of replica symmetry breaking on the NL has been discussed \cite{Nishimori2001,Nishimori2001b}. Additionally, an identity and an inequality between correlation functions have been proved to hold in the same form as in the case of the Edwards-Anderson model, indicating that those gauge-non-invariant correlation functions behave similarly to those in the Edwards-Anderson model.

The present theory applies also to the infinite-range model with all-to-all interactions. One simply replaces the discrete variables $\tau_{ij}$ by continuous $J_{ij}$ and the summation over $\tau_{ij}$ by the integral in Eq.~(\ref{eq:gaussian}), and rescales $J_{ij}$ by an appropriate power of $N$ to have a meaningful thermodynamic limit as usual in the case of uncorrelated disorder \cite{Nishimori2001,Sherrington1975}. The infinite-range model with uncorrelated disorder, the Sherrington-Kirkpatrick model \cite{Sherrington1975}, is known to have replica symmetry breaking \cite{Parisi1980} and temperature chaos \cite{Rizzo2003}, and therefore the anomalous properties described in the above paragraph apply to the present model with infinite-range interactions. Whether or not the same is true for finite-dimensional problems is an open question. A subtle point to note is that the detailed analyses of the Sherrington-Kirkpatrick model are carried out after the thermodynamic limit is taken as usual in the mean-field models whereas the thermodynamic limit is taken after all other manipulations in our case. Whether or not this aspect affects the relation between the present results and those for the Sherrington-Kirkpatrick model is a non-trivial problem for future studies.

The probability distribution $P(\tau)$ of Eq.~(\ref{eq:P_correlated}) is admittedly artificial and unlikely to mirror experimental conditions in random magnets or quantum devices. This distribution was deliberately crafted to simplify the mathematical analysis. However, we stress the importance of the very existence of such a model, which demonstrates that a spin glass model can be tuned to exhibit anomalous behavior through relatively straightforward constructions. In particular, the presence of support on a finite interval in the magnetization distribution function merits careful examination, as it directly challenges the conventional arguments against replica symmetry breaking on the NL in the Edwards-Anderson model \cite{Nishimori2001, Nishimori2001b}.  This is also important from a machine learning and statistical inference point of view, as described in detail in Ref.~\cite{Zdeborova2016}. Although it is improbable that the Edwards-Anderson model itself displays support on a finite interval in its magnetization distribution, our findings emphasize the need for more rigorous analyses that extend beyond traditional arguments.

\appendix*
\section{Properties of $P(\tau)$}
\label{sec:appendix}
This Appendix discusses some of the properties of the disorder distribution function $P(\tau)$ in Eq.~(\ref{eq:P_correlated}), in particular its behavior in the limit $\beta_p\to\infty$ and the possibility of generating samples of $\{\tau\}$ numerically.

\subsection{The limit $\beta_p\to\infty$}
First, we consider the limit $\beta_p\to\infty$, in which only the ground state spin configuration survives in the partition function $Z_{\tau}(\beta_p)$ in the denominator of $P(\tau)$.  Let us assume that the pure ferromagnetic state (all-spin up or all-spin down) with the energy $E=-\sum\tau_{ij}$ is a ground state for a given $\{\tau\}$. The probability distribution then approaches as $\beta_p\to\infty$ to
\begin{align}
    P(\tau)\to \frac{1}{A}\,\frac{e^{\beta_p\sum\tau_{ij}}}{c\,e^{\beta_p\sum\tau_{ij}}}=\frac{1}{cA}>0,
\end{align}
where $c$ accounts for degeneracy\footnote{Non pure-ferromagnetic spin states may have the same energy $-\sum\tau_{ij}$. For instance, if two consecutive horizontal bonds are antiferromagnetic while all other bonds are ferromagnetic, the spin enclosed by these two antiferromagnetic bonds can align either up or down without altering the system's energy.}.
Thus, the disorder configurations with the pure ferromagnetic state as (one of) the ground state(s) survive in this limit. Figure \ref{fig1} illustrates such an example, where pairs of adjacent frustrated plaquettes are located at a finite distance from other such pairs, leading to the pure ferromagnetic spin configuration as the ground state. There can be very many, possibly exponentially many, other non-trivial disorder configurations with the pure ferromagnetic state as a ground state.

It is useful to consider a disorder configuration whose ground state is not purely ferromagnetic, i.e., the pure ferromagnetic state is an excited state.   Then the ground-state energy $E_{\rm g}$ is lower than $-\sum\tau_{ij}$ by $-\Delta E(<0)$,
\begin{align}
    E_{\rm g}=-\sum\tau_{ij}-\Delta E.
\end{align}
The probability of this configuration in the limit $\beta_p\to\infty$ is, with $c$ being degeneracy,
\begin{align}
    P(\tau)\to \frac{1}{A}\,\frac{e^{\beta_p\sum\tau_{ij}}}{c\,e^{\beta_p(\sum\tau_{ij}+\Delta E)}}\to 0.
\end{align}
Consequently, these disorder configurations do not survive in this limit. We conclude that only the disorder configurations with the pure ferromagnetic state as a ground state have a finite probability in the limit $\beta_p\to \infty$.

\subsection{Numerical simulations}
Numerical simulations of the present model are not straightforward because it is non-trivial to generate the disorder distribution of Eq.~(\ref{eq:P_correlated}) with the partition function $Z_{\tau}(\beta_p)$ in the denominator. If we try to apply the Monte Carlo method to generate a sequence of $\{\tau\}$, we have to evaluate the ratio of $Z_{\tau'}(\beta_p)$ to $Z_{\tau}(\beta_p)$, or the difference of the free energy $\Delta F=F_{\tau'}(\beta_p)-F_{\tau}(\beta_p)$, for a bond flip trial $\tau\to\tau'$. But the bond variables $\{\tau\}$ are globally correlated as can be verified by the high-temperature (small-$\beta_p$) expansion of the free energy $F_{\tau}(\beta_p)$, which generates all types connected graphs, implying multi-body long-range interactions of $\{\tau\}$ variables. Thus, the evaluation of $\Delta F$ is quite time consuming.

Also, applying one of the various Monte Carlo methods to evaluate the free energy (see Refs.~\cite{Bi2015,Yasuda2022} and references therein) may not be particularly efficient in the present context, as it necessitates repeating the procedure very many times to generate a large number of samples, each of which involves a time-intensive free-energy calculation.

A viable alternative could involve utilizing the tensor-network method to evaluate the partition function, as demonstrated in Ref.~\cite{Zhu2023}, where the two-dimensional spin-glass problem is studied with a disorder probability distribution proportional to $Z_{\tau}(\beta_p)$\footnote{This example inspired the present author to try the probability distribution of Eq.~(\ref{eq:P_correlated}).}. It should be noted that the method requires adaptation to three or higher dimensions, given that the two-dimensional Edwards-Anderson model does not exhibit a spin glass phase at finite temperatures. However, such a generalization presents a non-trivial challenge. See, for example, Refs.~\cite{Okunishi2022,Liu2023}.


\begin{thebibliography}{103}%
\makeatletter
\providecommand \@ifxundefined [1]{%
 \@ifx{#1\undefined}
}%
\providecommand \@ifnum [1]{%
 \ifnum #1\expandafter \@firstoftwo
 \else \expandafter \@secondoftwo
 \fi
}%
\providecommand \@ifx [1]{%
 \ifx #1\expandafter \@firstoftwo
 \else \expandafter \@secondoftwo
 \fi
}%
\providecommand \natexlab [1]{#1}%
\providecommand \enquote  [1]{``#1''}%
\providecommand \bibnamefont  [1]{#1}%
\providecommand \bibfnamefont [1]{#1}%
\providecommand \citenamefont [1]{#1}%
\providecommand \href@noop [0]{\@secondoftwo}%
\providecommand \href [0]{\begingroup \@sanitize@url \@href}%
\providecommand \@href[1]{\@@startlink{#1}\@@href}%
\providecommand \@@href[1]{\endgroup#1\@@endlink}%
\providecommand \@sanitize@url [0]{\catcode `\\12\catcode `\$12\catcode `\&12\catcode `\#12\catcode `\^12\catcode `\_12\catcode `\%12\relax}%
\providecommand \@@startlink[1]{}%
\providecommand \@@endlink[0]{}%
\providecommand \url  [0]{\begingroup\@sanitize@url \@url }%
\providecommand \@url [1]{\endgroup\@href {#1}{\urlprefix }}%
\providecommand \urlprefix  [0]{URL }%
\providecommand \Eprint [0]{\href }%
\providecommand \doibase [0]{https://doi.org/}%
\providecommand \selectlanguage [0]{\@gobble}%
\providecommand \bibinfo  [0]{\@secondoftwo}%
\providecommand \bibfield  [0]{\@secondoftwo}%
\providecommand \translation [1]{[#1]}%
\providecommand \BibitemOpen [0]{}%
\providecommand \bibitemStop [0]{}%
\providecommand \bibitemNoStop [0]{.\EOS\space}%
\providecommand \EOS [0]{\spacefactor3000\relax}%
\providecommand \BibitemShut  [1]{\csname bibitem#1\endcsname}%
\let\auto@bib@innerbib\@empty
\bibitem [{\citenamefont {Nishimori}(2001)}]{Nishimori2001}%
  \BibitemOpen
  \bibfield  {author} {\bibinfo {author} {\bibfnamefont {H.}~\bibnamefont {Nishimori}},\ }\href {https://doi.org/10.1093/acprof:oso/9780198509417.001.0001} {\emph {\bibinfo {title} {{Statistical Physics of Spin Glasses and Information Processing: An Introduction}}}}\ (\bibinfo  {publisher} {Oxford University Press},\ \bibinfo {address} {Oxford},\ \bibinfo {year} {2001})\BibitemShut {NoStop}%
\bibitem [{\citenamefont {Huang}(2022)}]{Huang2022}%
  \BibitemOpen
  \bibfield  {author} {\bibinfo {author} {\bibfnamefont {H.}~\bibnamefont {Huang}},\ }\href {https://doi.org/10.1007/978-981-16-7570-6} {\emph {\bibinfo {title} {{Statistical Mechanics of Neural Networks}}}}\ (\bibinfo  {publisher} {Springer},\ \bibinfo {address} {Singapore},\ \bibinfo {year} {2022})\BibitemShut {NoStop}%
\bibitem [{\citenamefont {Charbonneau}\ \emph {et~al.}(2023)\citenamefont {Charbonneau}, \citenamefont {Marinari}, \citenamefont {Parisi}, \citenamefont {Ricci-tersenghi}, \citenamefont {Sicuro}, \citenamefont {Zamponi},\ and\ \citenamefont {M\'ezard}}]{Charbonneau2023}%
  \BibitemOpen
  \bibfield  {author} {\bibinfo {author} {\bibfnamefont {P.}~\bibnamefont {Charbonneau}}, \bibinfo {author} {\bibfnamefont {E.}~\bibnamefont {Marinari}}, \bibinfo {author} {\bibfnamefont {G.}~\bibnamefont {Parisi}}, \bibinfo {author} {\bibfnamefont {F.}~\bibnamefont {Ricci-tersenghi}}, \bibinfo {author} {\bibfnamefont {G.}~\bibnamefont {Sicuro}}, \bibinfo {author} {\bibfnamefont {F.}~\bibnamefont {Zamponi}},\ and\ \bibinfo {author} {\bibfnamefont {M.}~\bibnamefont {M\'ezard}},\ }\href {https://doi.org/10.1142/13341} {\emph {\bibinfo {title} {Spin Glass Theory and Far Beyond: Replica Symmetry Breaking after 40 Years}}}\ (\bibinfo  {publisher} {World Scientific},\ \bibinfo {address} {Singapore},\ \bibinfo {year} {2023})\BibitemShut {NoStop}%
\bibitem [{\citenamefont {Sherrington}\ and\ \citenamefont {Kirkpatrick}(1975)}]{Sherrington1975}%
  \BibitemOpen
  \bibfield  {author} {\bibinfo {author} {\bibfnamefont {D.}~\bibnamefont {Sherrington}}\ and\ \bibinfo {author} {\bibfnamefont {S.}~\bibnamefont {Kirkpatrick}},\ }\bibfield  {title} {\bibinfo {title} {{Solvable model of a spin glass}},\ }\href {https://doi.org/10.1103/PhysRevLett.35.1792} {\bibfield  {journal} {\bibinfo  {journal} {Phys. Rev. Lett.}\ }\textbf {\bibinfo {volume} {35}},\ \bibinfo {pages} {1792} (\bibinfo {year} {1975})}\BibitemShut {NoStop}%
\bibitem [{\citenamefont {Parisi}(1980)}]{Parisi1980}%
  \BibitemOpen
  \bibfield  {author} {\bibinfo {author} {\bibfnamefont {G.}~\bibnamefont {Parisi}},\ }\bibfield  {title} {\bibinfo {title} {A sequence of approximated solutions to the {SK} model for spin glasses},\ }\href {https://doi.org/10.1088/0305-4470/13/4/009} {\bibfield  {journal} {\bibinfo  {journal} {J. Phys. A}\ }\textbf {\bibinfo {volume} {13}},\ \bibinfo {pages} {L115} (\bibinfo {year} {1980})}\BibitemShut {NoStop}%
\bibitem [{\citenamefont {Nishimori}(1981)}]{Nishimori1981}%
  \BibitemOpen
  \bibfield  {author} {\bibinfo {author} {\bibfnamefont {H.}~\bibnamefont {Nishimori}},\ }\bibfield  {title} {\bibinfo {title} {Internal energy, specific heat and correlation function of the bond-random {Ising} model},\ }\href {https://doi.org/10.1143/PTP.66.1169} {\bibfield  {journal} {\bibinfo  {journal} {Prog. Theor. Phys.}\ }\textbf {\bibinfo {volume} {66}},\ \bibinfo {pages} {1169} (\bibinfo {year} {1981})}\BibitemShut {NoStop}%
\bibitem [{\citenamefont {Nishimori}(1980)}]{Nishimori1980}%
  \BibitemOpen
  \bibfield  {author} {\bibinfo {author} {\bibfnamefont {H.}~\bibnamefont {Nishimori}},\ }\bibfield  {title} {\bibinfo {title} {Exact results and critical properties of the {Ising} model with competing interactions},\ }\href {https://doi.org/10.1088/0022-3719/13/21/012} {\bibfield  {journal} {\bibinfo  {journal} {J. Phys. C}\ }\textbf {\bibinfo {volume} {13}},\ \bibinfo {pages} {4071} (\bibinfo {year} {1980})}\BibitemShut {NoStop}%
\bibitem [{\citenamefont {Edwards}\ and\ \citenamefont {Anderson}(1975)}]{Edwards1975}%
  \BibitemOpen
  \bibfield  {author} {\bibinfo {author} {\bibfnamefont {S.~F.}\ \bibnamefont {Edwards}}\ and\ \bibinfo {author} {\bibfnamefont {P.~W.}\ \bibnamefont {Anderson}},\ }\bibfield  {title} {\bibinfo {title} {Theory of spin glasses},\ }\href {https://doi.org/10.1088/0305-4608/5/5/017} {\bibfield  {journal} {\bibinfo  {journal} {J. Phys. F}\ }\textbf {\bibinfo {volume} {5}},\ \bibinfo {pages} {965} (\bibinfo {year} {1975})}\BibitemShut {NoStop}%
\bibitem [{\citenamefont {Nogueira}\ \emph {et~al.}(1999)\citenamefont {Nogueira}, \citenamefont {Coutinho}, \citenamefont {Nobre},\ and\ \citenamefont {Curado}}]{Nogueira1999}%
  \BibitemOpen
  \bibfield  {author} {\bibinfo {author} {\bibfnamefont {E.~J.}\ \bibnamefont {Nogueira}}, \bibinfo {author} {\bibfnamefont {S.}~\bibnamefont {Coutinho}}, \bibinfo {author} {\bibfnamefont {F.~D.}\ \bibnamefont {Nobre}},\ and\ \bibinfo {author} {\bibfnamefont {E.}~\bibnamefont {Curado}},\ }\bibfield  {title} {\bibinfo {title} {Universality in short-range {Ising} spin glasses},\ }\href {https://doi.org/https://doi.org/10.1016/S0378-4371(99)00229-0} {\bibfield  {journal} {\bibinfo  {journal} {Physica A}\ }\textbf {\bibinfo {volume} {271}},\ \bibinfo {pages} {125} (\bibinfo {year} {1999})}\BibitemShut {NoStop}%
\bibitem [{\citenamefont {Honecker}\ \emph {et~al.}(2001)\citenamefont {Honecker}, \citenamefont {Picco},\ and\ \citenamefont {Pujol}}]{Honecker2001}%
  \BibitemOpen
  \bibfield  {author} {\bibinfo {author} {\bibfnamefont {A.}~\bibnamefont {Honecker}}, \bibinfo {author} {\bibfnamefont {M.}~\bibnamefont {Picco}},\ and\ \bibinfo {author} {\bibfnamefont {P.}~\bibnamefont {Pujol}},\ }\bibfield  {title} {\bibinfo {title} {Universality class of the {Nishimori} point in the {2D $\pm J$} random-bond {Ising} model},\ }\href {https://doi.org/10.1103/PhysRevLett.87.047201} {\bibfield  {journal} {\bibinfo  {journal} {Phys. Rev. Lett.}\ }\textbf {\bibinfo {volume} {87}},\ \bibinfo {pages} {047201} (\bibinfo {year} {2001})}\BibitemShut {NoStop}%
\bibitem [{\citenamefont {J\"org}\ \emph {et~al.}(2006)\citenamefont {J\"org}, \citenamefont {Lukic}, \citenamefont {Marinari},\ and\ \citenamefont {Martin}}]{Jorg2006}%
  \BibitemOpen
  \bibfield  {author} {\bibinfo {author} {\bibfnamefont {T.}~\bibnamefont {J\"org}}, \bibinfo {author} {\bibfnamefont {J.}~\bibnamefont {Lukic}}, \bibinfo {author} {\bibfnamefont {E.}~\bibnamefont {Marinari}},\ and\ \bibinfo {author} {\bibfnamefont {O.~C.}\ \bibnamefont {Martin}},\ }\bibfield  {title} {\bibinfo {title} {Strong universality and algebraic scaling in two-dimensional {Ising} spin glasses},\ }\href {https://doi.org/10.1103/PhysRevLett.96.237205} {\bibfield  {journal} {\bibinfo  {journal} {Phys. Rev. Lett.}\ }\textbf {\bibinfo {volume} {96}},\ \bibinfo {pages} {237205} (\bibinfo {year} {2006})}\BibitemShut {NoStop}%
\bibitem [{\citenamefont {Hasenbusch}\ \emph {et~al.}(2008{\natexlab{a}})\citenamefont {Hasenbusch}, \citenamefont {Pelissetto},\ and\ \citenamefont {Vicari}}]{Hasenbusch2008}%
  \BibitemOpen
  \bibfield  {author} {\bibinfo {author} {\bibfnamefont {M.}~\bibnamefont {Hasenbusch}}, \bibinfo {author} {\bibfnamefont {A.}~\bibnamefont {Pelissetto}},\ and\ \bibinfo {author} {\bibfnamefont {E.}~\bibnamefont {Vicari}},\ }\bibfield  {title} {\bibinfo {title} {{Critical behavior of three-dimensional Ising spin glass models}},\ }\href {https://doi.org/10.1103/PhysRevB.78.214205} {\bibfield  {journal} {\bibinfo  {journal} {Phy. Rev. B}\ }\textbf {\bibinfo {volume} {78}},\ \bibinfo {pages} {214205} (\bibinfo {year} {2008}{\natexlab{a}})}\BibitemShut {NoStop}%
\bibitem [{\citenamefont {{Parisen Toldin}}\ \emph {et~al.}(2010)\citenamefont {{Parisen Toldin}}, \citenamefont {Pelissetto},\ and\ \citenamefont {Vicari}}]{ParisenToldin2010}%
  \BibitemOpen
  \bibfield  {author} {\bibinfo {author} {\bibfnamefont {F.}~\bibnamefont {{Parisen Toldin}}}, \bibinfo {author} {\bibfnamefont {A.}~\bibnamefont {Pelissetto}},\ and\ \bibinfo {author} {\bibfnamefont {E.}~\bibnamefont {Vicari}},\ }\bibfield  {title} {\bibinfo {title} {{Universality of the glassy transitions in the two-dimensional $\pm J$ Ising model}},\ }\href {https://doi.org/10.1103/PhysRevE.82.021106} {\bibfield  {journal} {\bibinfo  {journal} {Phys. Rev. E}\ }\textbf {\bibinfo {volume} {82}},\ \bibinfo {pages} {1} (\bibinfo {year} {2010})}\BibitemShut {NoStop}%
\bibitem [{\citenamefont {Katzgraber}\ \emph {et~al.}(2006)\citenamefont {Katzgraber}, \citenamefont {K\"orner},\ and\ \citenamefont {Young}}]{Katzgraber2006}%
  \BibitemOpen
  \bibfield  {author} {\bibinfo {author} {\bibfnamefont {H.~G.}\ \bibnamefont {Katzgraber}}, \bibinfo {author} {\bibfnamefont {M.}~\bibnamefont {K\"orner}},\ and\ \bibinfo {author} {\bibfnamefont {A.~P.}\ \bibnamefont {Young}},\ }\bibfield  {title} {\bibinfo {title} {{Universality in three-dimensional Ising spin glasses: A Monte Carlo study}},\ }\href {https://doi.org/10.1103/PhysRevB.73.224432} {\bibfield  {journal} {\bibinfo  {journal} {Phys. Rev. B}\ }\textbf {\bibinfo {volume} {73}},\ \bibinfo {pages} {224432} (\bibinfo {year} {2006})}\BibitemShut {NoStop}%
\bibitem [{\citenamefont {Katzgraber}\ \emph {et~al.}(2007)\citenamefont {Katzgraber}, \citenamefont {Lee},\ and\ \citenamefont {Campbell}}]{Katzgraber2007}%
  \BibitemOpen
  \bibfield  {author} {\bibinfo {author} {\bibfnamefont {H.~G.}\ \bibnamefont {Katzgraber}}, \bibinfo {author} {\bibfnamefont {L.~W.}\ \bibnamefont {Lee}},\ and\ \bibinfo {author} {\bibfnamefont {I.~A.}\ \bibnamefont {Campbell}},\ }\bibfield  {title} {\bibinfo {title} {Effective critical behavior of the two-dimensional {Ising} spin glass with bimodal interactions},\ }\href {https://doi.org/10.1103/PhysRevB.75.014412} {\bibfield  {journal} {\bibinfo  {journal} {Phys. Rev. B}\ }\textbf {\bibinfo {volume} {75}},\ \bibinfo {pages} {014412} (\bibinfo {year} {2007})}\BibitemShut {NoStop}%
\bibitem [{\citenamefont {Hasenbusch}\ \emph {et~al.}(2008{\natexlab{b}})\citenamefont {Hasenbusch}, \citenamefont {Pelissetto},\ and\ \citenamefont {Vicari}}]{Hasenbusch2008a}%
  \BibitemOpen
  \bibfield  {author} {\bibinfo {author} {\bibfnamefont {M.}~\bibnamefont {Hasenbusch}}, \bibinfo {author} {\bibfnamefont {A.}~\bibnamefont {Pelissetto}},\ and\ \bibinfo {author} {\bibfnamefont {E.}~\bibnamefont {Vicari}},\ }\bibfield  {title} {\bibinfo {title} {{The critical behavior of 3D Ising spin glass models: Universality and scaling corrections}},\ }\href {https://doi.org/10.1088/1742-5468/2008/02/L02001} {\bibfield  {journal} {\bibinfo  {journal} {J. Stat. Mech.}\ }\textbf {\bibinfo {volume} {2008}},\ \bibinfo {pages} {L02001} (\bibinfo {year} {2008}{\natexlab{b}})}\BibitemShut {NoStop}%
\bibitem [{\citenamefont {Mydosh}(1993)}]{Mydosh1993}%
  \BibitemOpen
  \bibfield  {author} {\bibinfo {author} {\bibfnamefont {J.~A.}\ \bibnamefont {Mydosh}},\ }\href {https://doi.org/10.1201/9781482295191} {\emph {\bibinfo {title} {Spin glasses: An experimental introduction}}}\ (\bibinfo  {publisher} {CRC Press},\ \bibinfo {address} {London},\ \bibinfo {year} {1993})\BibitemShut {NoStop}%
\bibitem [{\citenamefont {Dennis}\ \emph {et~al.}(2002)\citenamefont {Dennis}, \citenamefont {Kitaev}, \citenamefont {Landahl},\ and\ \citenamefont {Preskill}}]{Dennis2002}%
  \BibitemOpen
  \bibfield  {author} {\bibinfo {author} {\bibfnamefont {E.}~\bibnamefont {Dennis}}, \bibinfo {author} {\bibfnamefont {A.}~\bibnamefont {Kitaev}}, \bibinfo {author} {\bibfnamefont {A.}~\bibnamefont {Landahl}},\ and\ \bibinfo {author} {\bibfnamefont {J.}~\bibnamefont {Preskill}},\ }\bibfield  {title} {\bibinfo {title} {{Topological quantum memory}},\ }\href {https://doi.org/10.1063/1.1499754} {\bibfield  {journal} {\bibinfo  {journal} {J. Math. Phys.}\ }\textbf {\bibinfo {volume} {43}},\ \bibinfo {pages} {4452} (\bibinfo {year} {2002})}\BibitemShut {NoStop}%
\bibitem [{\citenamefont {Wang}\ \emph {et~al.}(2002)\citenamefont {Wang}, \citenamefont {Harrington},\ and\ \citenamefont {Preskill}}]{Wang2002}%
  \BibitemOpen
  \bibfield  {author} {\bibinfo {author} {\bibfnamefont {C.}~\bibnamefont {Wang}}, \bibinfo {author} {\bibfnamefont {J.}~\bibnamefont {Harrington}},\ and\ \bibinfo {author} {\bibfnamefont {J.}~\bibnamefont {Preskill}},\ }\bibfield  {title} {\bibinfo {title} {{Confinement-Higgs transition in a disordered gauge theory and the accuracy threshold for quantum memory}},\ }\href {https://doi.org/10.1016/S0003-4916(02)00019-2} {\bibfield  {journal} {\bibinfo  {journal} {Ann. Phys.}\ }\textbf {\bibinfo {volume} {303}},\ \bibinfo {pages} {31} (\bibinfo {year} {2002})}\BibitemShut {NoStop}%
\bibitem [{\citenamefont {Bombin}\ \emph {et~al.}(2012)\citenamefont {Bombin}, \citenamefont {Andrist}, \citenamefont {Ohzeki}, \citenamefont {Katzgraber},\ and\ \citenamefont {Fisica}}]{Bombin2012}%
  \BibitemOpen
  \bibfield  {author} {\bibinfo {author} {\bibfnamefont {H.}~\bibnamefont {Bombin}}, \bibinfo {author} {\bibfnamefont {R.~S.}\ \bibnamefont {Andrist}}, \bibinfo {author} {\bibfnamefont {M.}~\bibnamefont {Ohzeki}}, \bibinfo {author} {\bibfnamefont {H.~G.}\ \bibnamefont {Katzgraber}},\ and\ \bibinfo {author} {\bibfnamefont {D.}~\bibnamefont {Fisica}},\ }\bibfield  {title} {\bibinfo {title} {Strong resilience of topological codes to depolarization},\ }\href {https://doi.org/10.1103/PhysRevX.2.021004Subject} {\bibfield  {journal} {\bibinfo  {journal} {Phys. Rev. X}\ }\textbf {\bibinfo {volume} {2}},\ \bibinfo {pages} {021004} (\bibinfo {year} {2012})}\BibitemShut {NoStop}%
\bibitem [{\citenamefont {Ohzeki}(2012)}]{Ohzeki2012}%
  \BibitemOpen
  \bibfield  {author} {\bibinfo {author} {\bibfnamefont {M.}~\bibnamefont {Ohzeki}},\ }\bibfield  {title} {\bibinfo {title} {Error threshold estimates for surface code with loss of qubits},\ }\href {https://doi.org/10.1103/PhysRevA.85.060301} {\bibfield  {journal} {\bibinfo  {journal} {Phys. Rev. A}\ }\textbf {\bibinfo {volume} {85}},\ \bibinfo {pages} {060301} (\bibinfo {year} {2012})}\BibitemShut {NoStop}%
\bibitem [{\citenamefont {Andrist}\ \emph {et~al.}(2015)\citenamefont {Andrist}, \citenamefont {Wootton},\ and\ \citenamefont {Katzgraber}}]{Andrist2015}%
  \BibitemOpen
  \bibfield  {author} {\bibinfo {author} {\bibfnamefont {R.~S.}\ \bibnamefont {Andrist}}, \bibinfo {author} {\bibfnamefont {J.~R.}\ \bibnamefont {Wootton}},\ and\ \bibinfo {author} {\bibfnamefont {H.~G.}\ \bibnamefont {Katzgraber}},\ }\bibfield  {title} {\bibinfo {title} {Error thresholds for abelian quantum double models: {Increasing} the bit-flip stability of topological quantum memory},\ }\href {https://doi.org/10.1103/PhysRevA.91.042331} {\bibfield  {journal} {\bibinfo  {journal} {Phys. Rev. A}\ }\textbf {\bibinfo {volume} {91}},\ \bibinfo {pages} {042331} (\bibinfo {year} {2015})}\BibitemShut {NoStop}%
\bibitem [{\citenamefont {Fujii}(2015)}]{fujii2015}%
  \BibitemOpen
  \bibfield  {author} {\bibinfo {author} {\bibfnamefont {K.}~\bibnamefont {Fujii}},\ }\href@noop {} {\emph {\bibinfo {title} {Quantum Computation with Topological Codes: from qubit to topological fault-tolerance}}}\ (\bibinfo  {publisher} {Springer},\ \bibinfo {year} {2015})\BibitemShut {NoStop}%
\bibitem [{\citenamefont {Andrist}\ \emph {et~al.}(2016)\citenamefont {Andrist}, \citenamefont {Katzgraber}, \citenamefont {Bombin},\ and\ \citenamefont {Martin-Delgado}}]{Andrist2016}%
  \BibitemOpen
  \bibfield  {author} {\bibinfo {author} {\bibfnamefont {R.~S.}\ \bibnamefont {Andrist}}, \bibinfo {author} {\bibfnamefont {H.~G.}\ \bibnamefont {Katzgraber}}, \bibinfo {author} {\bibfnamefont {H.}~\bibnamefont {Bombin}},\ and\ \bibinfo {author} {\bibfnamefont {M.~A.}\ \bibnamefont {Martin-Delgado}},\ }\bibfield  {title} {\bibinfo {title} {{Error tolerance of topological codes with independent bit-flip and measurement errors}},\ }\href {https://doi.org/10.1103/PhysRevA.94.012318} {\bibfield  {journal} {\bibinfo  {journal} {Phys. Rev. A}\ }\textbf {\bibinfo {volume} {012318}},\ \bibinfo {pages} {1} (\bibinfo {year} {2016})}\BibitemShut {NoStop}%
\bibitem [{\citenamefont {Kubica}\ \emph {et~al.}(2018)\citenamefont {Kubica}, \citenamefont {Beverland}, \citenamefont {Brand\~ao}, \citenamefont {Preskill},\ and\ \citenamefont {Svore}}]{Kubina2018}%
  \BibitemOpen
  \bibfield  {author} {\bibinfo {author} {\bibfnamefont {A.}~\bibnamefont {Kubica}}, \bibinfo {author} {\bibfnamefont {M.~E.}\ \bibnamefont {Beverland}}, \bibinfo {author} {\bibfnamefont {F.}~\bibnamefont {Brand\~ao}}, \bibinfo {author} {\bibfnamefont {J.}~\bibnamefont {Preskill}},\ and\ \bibinfo {author} {\bibfnamefont {K.~M.}\ \bibnamefont {Svore}},\ }\bibfield  {title} {\bibinfo {title} {Three-dimensional color code thresholds via statistical-mechanical mapping},\ }\href {https://doi.org/10.1103/PhysRevLett.120.180501} {\bibfield  {journal} {\bibinfo  {journal} {Phys. Rev. Lett.}\ }\textbf {\bibinfo {volume} {120}},\ \bibinfo {pages} {180501} (\bibinfo {year} {2018})}\BibitemShut {NoStop}%
\bibitem [{\citenamefont {Lee}\ \emph {et~al.}(2023)\citenamefont {Lee}, \citenamefont {Jian},\ and\ \citenamefont {Xu}}]{Lee2023}%
  \BibitemOpen
  \bibfield  {author} {\bibinfo {author} {\bibfnamefont {J.~Y.}\ \bibnamefont {Lee}}, \bibinfo {author} {\bibfnamefont {C.-M.}\ \bibnamefont {Jian}},\ and\ \bibinfo {author} {\bibfnamefont {C.}~\bibnamefont {Xu}},\ }\bibfield  {title} {\bibinfo {title} {Quantum criticality under decoherence or weak measurement},\ }\href {https://doi.org/10.1103/PRXQuantum.4.030317} {\bibfield  {journal} {\bibinfo  {journal} {PRX Quantum}\ }\textbf {\bibinfo {volume} {4}},\ \bibinfo {pages} {030317} (\bibinfo {year} {2023})}\BibitemShut {NoStop}%
\bibitem [{\citenamefont {Venn}\ \emph {et~al.}(2023)\citenamefont {Venn}, \citenamefont {Behrends},\ and\ \citenamefont {B\'eri}}]{Florian2023}%
  \BibitemOpen
  \bibfield  {author} {\bibinfo {author} {\bibfnamefont {F.}~\bibnamefont {Venn}}, \bibinfo {author} {\bibfnamefont {J.}~\bibnamefont {Behrends}},\ and\ \bibinfo {author} {\bibfnamefont {B.}~\bibnamefont {B\'eri}},\ }\bibfield  {title} {\bibinfo {title} {Coherent-error threshold for surface codes from {Majorana} delocalization},\ }\href {https://doi.org/10.1103/PhysRevLett.131.060603} {\bibfield  {journal} {\bibinfo  {journal} {Phys. Rev. Lett.}\ }\textbf {\bibinfo {volume} {131}},\ \bibinfo {pages} {060603} (\bibinfo {year} {2023})}\BibitemShut {NoStop}%
\bibitem [{\citenamefont {Fan}\ \emph {et~al.}(2024)\citenamefont {Fan}, \citenamefont {Bao}, \citenamefont {Altman},\ and\ \citenamefont {Vishwanath}}]{Fan2024}%
  \BibitemOpen
  \bibfield  {author} {\bibinfo {author} {\bibfnamefont {R.}~\bibnamefont {Fan}}, \bibinfo {author} {\bibfnamefont {Y.}~\bibnamefont {Bao}}, \bibinfo {author} {\bibfnamefont {E.}~\bibnamefont {Altman}},\ and\ \bibinfo {author} {\bibfnamefont {A.}~\bibnamefont {Vishwanath}},\ }\bibfield  {title} {\bibinfo {title} {Diagnostics of mixed-state topological order and breakdown of quantum memory},\ }\href {https://doi.org/10.1103/PRXQuantum.5.020343} {\bibfield  {journal} {\bibinfo  {journal} {PRX Quantum}\ }\textbf {\bibinfo {volume} {5}},\ \bibinfo {pages} {020343} (\bibinfo {year} {2024})}\BibitemShut {NoStop}%
\bibitem [{\citenamefont {Chen}\ and\ \citenamefont {Grover}(2024{\natexlab{a}})}]{Chen2024}%
  \BibitemOpen
  \bibfield  {author} {\bibinfo {author} {\bibfnamefont {Y.-H.}\ \bibnamefont {Chen}}\ and\ \bibinfo {author} {\bibfnamefont {T.}~\bibnamefont {Grover}},\ }\bibfield  {title} {\bibinfo {title} {Separability transitions in topological states induced by local decoherence},\ }\href {https://doi.org/10.1103/PhysRevLett.132.170602} {\bibfield  {journal} {\bibinfo  {journal} {Phys. Rev. Lett.}\ }\textbf {\bibinfo {volume} {132}},\ \bibinfo {pages} {170602} (\bibinfo {year} {2024}{\natexlab{a}})}\BibitemShut {NoStop}%
\bibitem [{\citenamefont {Chen}\ and\ \citenamefont {Grover}(2024{\natexlab{b}})}]{Chen2024a}%
  \BibitemOpen
  \bibfield  {author} {\bibinfo {author} {\bibfnamefont {Y.-H.}\ \bibnamefont {Chen}}\ and\ \bibinfo {author} {\bibfnamefont {T.}~\bibnamefont {Grover}},\ }\bibfield  {title} {\bibinfo {title} {Symmetry-enforced many-body separability transitions},\ }\href {https://doi.org/10.1103/PRXQuantum.5.030310} {\bibfield  {journal} {\bibinfo  {journal} {PRX Quantum}\ }\textbf {\bibinfo {volume} {5}},\ \bibinfo {pages} {030310} (\bibinfo {year} {2024}{\natexlab{b}})}\BibitemShut {NoStop}%
\bibitem [{\citenamefont {Dua}\ \emph {et~al.}(2024)\citenamefont {Dua}, \citenamefont {Kubica}, \citenamefont {Jiang}, \citenamefont {Flammia},\ and\ \citenamefont {Gullans}}]{Dua2024}%
  \BibitemOpen
  \bibfield  {author} {\bibinfo {author} {\bibfnamefont {A.}~\bibnamefont {Dua}}, \bibinfo {author} {\bibfnamefont {A.}~\bibnamefont {Kubica}}, \bibinfo {author} {\bibfnamefont {L.}~\bibnamefont {Jiang}}, \bibinfo {author} {\bibfnamefont {S.~T.}\ \bibnamefont {Flammia}},\ and\ \bibinfo {author} {\bibfnamefont {M.~J.}\ \bibnamefont {Gullans}},\ }\bibfield  {title} {\bibinfo {title} {Clifford-deformed surface codes},\ }\href {https://doi.org/10.1103/PRXQuantum.5.010347} {\bibfield  {journal} {\bibinfo  {journal} {PRX Quantum}\ }\textbf {\bibinfo {volume} {5}},\ \bibinfo {pages} {010347} (\bibinfo {year} {2024})}\BibitemShut {NoStop}%
\bibitem [{\citenamefont {Behrends}\ \emph {et~al.}(2024)\citenamefont {Behrends}, \citenamefont {Venn},\ and\ \citenamefont {B\'eri}}]{Behrends2024}%
  \BibitemOpen
  \bibfield  {author} {\bibinfo {author} {\bibfnamefont {J.}~\bibnamefont {Behrends}}, \bibinfo {author} {\bibfnamefont {F.}~\bibnamefont {Venn}},\ and\ \bibinfo {author} {\bibfnamefont {B.}~\bibnamefont {B\'eri}},\ }\bibfield  {title} {\bibinfo {title} {Surface codes, quantum circuits, and entanglement phases},\ }\href {https://doi.org/10.1103/PhysRevResearch.6.013137} {\bibfield  {journal} {\bibinfo  {journal} {Phys. Rev. Res.}\ }\textbf {\bibinfo {volume} {6}},\ \bibinfo {pages} {013137} (\bibinfo {year} {2024})}\BibitemShut {NoStop}%
\bibitem [{\citenamefont {Takada}\ \emph {et~al.}(2024)\citenamefont {Takada}, \citenamefont {Takeuchi},\ and\ \citenamefont {Fujii}}]{Takada2024}%
  \BibitemOpen
  \bibfield  {author} {\bibinfo {author} {\bibfnamefont {Y.}~\bibnamefont {Takada}}, \bibinfo {author} {\bibfnamefont {Y.}~\bibnamefont {Takeuchi}},\ and\ \bibinfo {author} {\bibfnamefont {K.}~\bibnamefont {Fujii}},\ }\bibfield  {title} {\bibinfo {title} {Ising model formulation for highly accurate topological color codes decoding},\ }\href {https://doi.org/10.1103/PhysRevResearch.6.013092} {\bibfield  {journal} {\bibinfo  {journal} {Phys. Rev. Res.}\ }\textbf {\bibinfo {volume} {6}},\ \bibinfo {pages} {013092} (\bibinfo {year} {2024})}\BibitemShut {NoStop}%
\bibitem [{\citenamefont {Hauser}\ \emph {et~al.}(2024)\citenamefont {Hauser}, \citenamefont {Bao}, \citenamefont {Sang}, \citenamefont {Lavasani}, \citenamefont {Agrawal},\ and\ \citenamefont {Fisher}}]{hauser2024information}%
  \BibitemOpen
  \bibfield  {author} {\bibinfo {author} {\bibfnamefont {J.}~\bibnamefont {Hauser}}, \bibinfo {author} {\bibfnamefont {Y.}~\bibnamefont {Bao}}, \bibinfo {author} {\bibfnamefont {S.}~\bibnamefont {Sang}}, \bibinfo {author} {\bibfnamefont {A.}~\bibnamefont {Lavasani}}, \bibinfo {author} {\bibfnamefont {U.}~\bibnamefont {Agrawal}},\ and\ \bibinfo {author} {\bibfnamefont {M.}~\bibnamefont {Fisher}},\ }\bibfield  {title} {\bibinfo {title} {Information dynamics in decohered quantum memory with repeated syndrome measurements: {A} dual approach},\ }\href {https://arxiv.org/abs/2407.07882} {\bibfield  {journal} {\bibinfo  {journal} {arXiv:2407.07882}\ } (\bibinfo {year} {2024})}\BibitemShut {NoStop}%
\bibitem [{\citenamefont {Su}\ \emph {et~al.}(2024)\citenamefont {Su}, \citenamefont {Yang},\ and\ \citenamefont {Jian}}]{Su2024}%
  \BibitemOpen
  \bibfield  {author} {\bibinfo {author} {\bibfnamefont {K.}~\bibnamefont {Su}}, \bibinfo {author} {\bibfnamefont {Z.}~\bibnamefont {Yang}},\ and\ \bibinfo {author} {\bibfnamefont {C.-M.}\ \bibnamefont {Jian}},\ }\bibfield  {title} {\bibinfo {title} {Tapestry of dualities in decohered quantum error correction codes},\ }\href {https://arxiv.org/abs/2401.17359} {\bibfield  {journal} {\bibinfo  {journal} {arxiv:2401.17359}\ } (\bibinfo {year} {2024})}\BibitemShut {NoStop}%
\bibitem [{\citenamefont {Zhang}\ \emph {et~al.}(2024)\citenamefont {Zhang}, \citenamefont {Agrawal},\ and\ \citenamefont {Vijay}}]{Zhang2024}%
  \BibitemOpen
  \bibfield  {author} {\bibinfo {author} {\bibfnamefont {Z.}~\bibnamefont {Zhang}}, \bibinfo {author} {\bibfnamefont {U.}~\bibnamefont {Agrawal}},\ and\ \bibinfo {author} {\bibfnamefont {S.}~\bibnamefont {Vijay}},\ }\bibfield  {title} {\bibinfo {title} {Quantum communication and mixed-state order in decohered symmetry-protected topological states},\ }\href {https://arxiv.org/abs/2405.05965} {\bibfield  {journal} {\bibinfo  {journal} {arXiv:2405.05965}\ } (\bibinfo {year} {2024})}\BibitemShut {NoStop}%
\bibitem [{\citenamefont {Lee}(2024)}]{lee2024}%
  \BibitemOpen
  \bibfield  {author} {\bibinfo {author} {\bibfnamefont {J.~Y.}\ \bibnamefont {Lee}},\ }\bibfield  {title} {\bibinfo {title} {Exact calculations of coherent information for toric codes under decoherence: {Identifying} the fundamental error threshold},\ }\href {https://arxiv.org/abs/2402.16937} {\bibfield  {journal} {\bibinfo  {journal} {arxiv:2402.16937}\ } (\bibinfo {year} {2024})}\BibitemShut {NoStop}%
\bibitem [{\citenamefont {Xiao}\ \emph {et~al.}(2024)\citenamefont {Xiao}, \citenamefont {Srivastava},\ and\ \citenamefont {Granath}}]{xiao2024}%
  \BibitemOpen
  \bibfield  {author} {\bibinfo {author} {\bibfnamefont {Y.}~\bibnamefont {Xiao}}, \bibinfo {author} {\bibfnamefont {B.}~\bibnamefont {Srivastava}},\ and\ \bibinfo {author} {\bibfnamefont {M.}~\bibnamefont {Granath}},\ }\bibfield  {title} {\bibinfo {title} {Exact results on finite size corrections for surface codes tailored to biased noise},\ }\href {https://arxiv.org/abs/2401.04008} {\bibfield  {journal} {\bibinfo  {journal} {arxiv:2401.04008}\ } (\bibinfo {year} {2024})}\BibitemShut {NoStop}%
\bibitem [{\citenamefont {Li}\ \emph {et~al.}(2024)\citenamefont {Li}, \citenamefont {O'Dea},\ and\ \citenamefont {Khemani}}]{li2024}%
  \BibitemOpen
  \bibfield  {author} {\bibinfo {author} {\bibfnamefont {Y.}~\bibnamefont {Li}}, \bibinfo {author} {\bibfnamefont {N.}~\bibnamefont {O'Dea}},\ and\ \bibinfo {author} {\bibfnamefont {V.}~\bibnamefont {Khemani}},\ }\bibfield  {title} {\bibinfo {title} {Perturbative stability and error correction thresholds of quantum codes},\ }\href {https://arxiv.org/abs/2406.15757} {\bibfield  {journal} {\bibinfo  {journal} {arxiv:2406.15757}\ } (\bibinfo {year} {2024})}\BibitemShut {NoStop}%
\bibitem [{\citenamefont {Wang}\ \emph {et~al.}(2024)\citenamefont {Wang}, \citenamefont {Song}, \citenamefont {Meng},\ and\ \citenamefont {Grover}}]{wang2024a}%
  \BibitemOpen
  \bibfield  {author} {\bibinfo {author} {\bibfnamefont {T.-T.}\ \bibnamefont {Wang}}, \bibinfo {author} {\bibfnamefont {M.}~\bibnamefont {Song}}, \bibinfo {author} {\bibfnamefont {Z.~Y.}\ \bibnamefont {Meng}},\ and\ \bibinfo {author} {\bibfnamefont {T.}~\bibnamefont {Grover}},\ }\bibfield  {title} {\bibinfo {title} {An analog of topological entanglement entropy for mixed states},\ }\href {https://arxiv.org/abs/2407.20500} {\bibfield  {journal} {\bibinfo  {journal} {arxiv:2407.20500}\ } (\bibinfo {year} {2024})}\BibitemShut {NoStop}%
\bibitem [{\citenamefont {Niwa}\ and\ \citenamefont {Lee}(2024)}]{niwa2024}%
  \BibitemOpen
  \bibfield  {author} {\bibinfo {author} {\bibfnamefont {R.}~\bibnamefont {Niwa}}\ and\ \bibinfo {author} {\bibfnamefont {J.~Y.}\ \bibnamefont {Lee}},\ }\bibfield  {title} {\bibinfo {title} {Coherent information for {CSS} codes under decoherence},\ }\href {https://arxiv.org/abs/2407.02564} {\bibfield  {journal} {\bibinfo  {journal} {arxiv:2407.02564}\ } (\bibinfo {year} {2024})}\BibitemShut {NoStop}%
\bibitem [{\citenamefont {Clemens}\ \emph {et~al.}(2004)\citenamefont {Clemens}, \citenamefont {Siddiqui},\ and\ \citenamefont {Gea-Banacloche}}]{Clemens2004}%
  \BibitemOpen
  \bibfield  {author} {\bibinfo {author} {\bibfnamefont {J.~P.}\ \bibnamefont {Clemens}}, \bibinfo {author} {\bibfnamefont {S.}~\bibnamefont {Siddiqui}},\ and\ \bibinfo {author} {\bibfnamefont {J.}~\bibnamefont {Gea-Banacloche}},\ }\bibfield  {title} {\bibinfo {title} {Quantum error correction against correlated noise},\ }\href {https://doi.org/10.1103/PhysRevA.69.062313} {\bibfield  {journal} {\bibinfo  {journal} {Phys. Rev. A}\ }\textbf {\bibinfo {volume} {69}},\ \bibinfo {pages} {062313} (\bibinfo {year} {2004})}\BibitemShut {NoStop}%
\bibitem [{\citenamefont {Klesse}\ and\ \citenamefont {Frank}(2005)}]{Klesse2005}%
  \BibitemOpen
  \bibfield  {author} {\bibinfo {author} {\bibfnamefont {R.}~\bibnamefont {Klesse}}\ and\ \bibinfo {author} {\bibfnamefont {S.}~\bibnamefont {Frank}},\ }\bibfield  {title} {\bibinfo {title} {Quantum error correction in spatially correlated quantum noise},\ }\href {https://doi.org/10.1103/PhysRevLett.95.230503} {\bibfield  {journal} {\bibinfo  {journal} {Phys. Rev. Lett.}\ }\textbf {\bibinfo {volume} {95}},\ \bibinfo {pages} {230503} (\bibinfo {year} {2005})}\BibitemShut {NoStop}%
\bibitem [{\citenamefont {Aharonov}\ and\ \citenamefont {Ben-Or}(2008)}]{Aharonov2008}%
  \BibitemOpen
  \bibfield  {author} {\bibinfo {author} {\bibfnamefont {D.}~\bibnamefont {Aharonov}}\ and\ \bibinfo {author} {\bibfnamefont {M.}~\bibnamefont {Ben-Or}},\ }\bibfield  {title} {\bibinfo {title} {Fault-tolerant quantum computation with constant error rate},\ }\href {https://doi.org/10.1137/S0097539799359385} {\bibfield  {journal} {\bibinfo  {journal} {SIAM J. Comp.}\ }\textbf {\bibinfo {volume} {38}},\ \bibinfo {pages} {1207} (\bibinfo {year} {2008})}\BibitemShut {NoStop}%
\bibitem [{\citenamefont {Preskill}(2013)}]{Preskill2013}%
  \BibitemOpen
  \bibfield  {author} {\bibinfo {author} {\bibfnamefont {J.}~\bibnamefont {Preskill}},\ }\bibfield  {title} {\bibinfo {title} {Sufficient condition on noise correlations for scalable quantum computing},\ }\href {https://doi.org/10.26421/QIC13.3-4-1} {\bibfield  {journal} {\bibinfo  {journal} {Quantum Inf. Comput.}\ }\textbf {\bibinfo {volume} {13}},\ \bibinfo {pages} {181} (\bibinfo {year} {2013})}\BibitemShut {NoStop}%
\bibitem [{\citenamefont {Ball}\ \emph {et~al.}(2016)\citenamefont {Ball}, \citenamefont {Stace}, \citenamefont {Flammia},\ and\ \citenamefont {Biercuk}}]{Ball2016}%
  \BibitemOpen
  \bibfield  {author} {\bibinfo {author} {\bibfnamefont {H.}~\bibnamefont {Ball}}, \bibinfo {author} {\bibfnamefont {T.~M.}\ \bibnamefont {Stace}}, \bibinfo {author} {\bibfnamefont {S.~T.}\ \bibnamefont {Flammia}},\ and\ \bibinfo {author} {\bibfnamefont {M.~J.}\ \bibnamefont {Biercuk}},\ }\bibfield  {title} {\bibinfo {title} {Effect of noise correlations on randomized benchmarking},\ }\href {https://doi.org/10.1103/PhysRevA.93.022303} {\bibfield  {journal} {\bibinfo  {journal} {Phys. Rev. A}\ }\textbf {\bibinfo {volume} {93}},\ \bibinfo {pages} {022303} (\bibinfo {year} {2016})}\BibitemShut {NoStop}%
\bibitem [{\citenamefont {Wilen}\ \emph {et~al.}(2021)\citenamefont {Wilen}, \citenamefont {Abdullah}, \citenamefont {Kurinsky}, \citenamefont {Stanford}, \citenamefont {Cardani}, \citenamefont {D'Imperio}, \citenamefont {Tomei}, \citenamefont {Faoro}, \citenamefont {Ioffe}, \citenamefont {Liu}, \citenamefont {Opremcak}, \citenamefont {Christensen}, \citenamefont {DuBois},\ and\ \citenamefont {McDermott}}]{Wilen2021}%
  \BibitemOpen
  \bibfield  {author} {\bibinfo {author} {\bibfnamefont {C.~D.}\ \bibnamefont {Wilen}}, \bibinfo {author} {\bibfnamefont {S.}~\bibnamefont {Abdullah}}, \bibinfo {author} {\bibfnamefont {N.~A.}\ \bibnamefont {Kurinsky}}, \bibinfo {author} {\bibfnamefont {C.}~\bibnamefont {Stanford}}, \bibinfo {author} {\bibfnamefont {L.}~\bibnamefont {Cardani}}, \bibinfo {author} {\bibfnamefont {G.}~\bibnamefont {D'Imperio}}, \bibinfo {author} {\bibfnamefont {C.}~\bibnamefont {Tomei}}, \bibinfo {author} {\bibfnamefont {L.}~\bibnamefont {Faoro}}, \bibinfo {author} {\bibfnamefont {L.~B.}\ \bibnamefont {Ioffe}}, \bibinfo {author} {\bibfnamefont {C.~H.}\ \bibnamefont {Liu}}, \bibinfo {author} {\bibfnamefont {A.}~\bibnamefont {Opremcak}}, \bibinfo {author} {\bibfnamefont {B.~G.}\ \bibnamefont {Christensen}}, \bibinfo {author} {\bibfnamefont {J.~L.}\ \bibnamefont {DuBois}},\ and\ \bibinfo {author} {\bibfnamefont {R.}~\bibnamefont {McDermott}},\ }\bibfield  {title} {\bibinfo {title} {{Correlated charge noise and relaxation errors in
  superconducting qubits}},\ }\href {https://doi.org/10.1038/s41586-021-03557-5} {\bibfield  {journal} {\bibinfo  {journal} {Nature}\ }\textbf {\bibinfo {volume} {594}},\ \bibinfo {pages} {369} (\bibinfo {year} {2021})}\BibitemShut {NoStop}%
\bibitem [{\citenamefont {Chubb}\ and\ \citenamefont {Flammia}(2021)}]{Chubb2021}%
  \BibitemOpen
  \bibfield  {author} {\bibinfo {author} {\bibfnamefont {C.~T.}\ \bibnamefont {Chubb}}\ and\ \bibinfo {author} {\bibfnamefont {S.~T.}\ \bibnamefont {Flammia}},\ }\bibfield  {title} {\bibinfo {title} {Statistical mechanical models for quantum codes with correlated noise},\ }\href {https://doi.org/10.4171/aihpd/105} {\bibfield  {journal} {\bibinfo  {journal} {Ann. Inst. Henri Poincar\'e Comb. Phys. Interact.}\ }\textbf {\bibinfo {volume} {8}} (\bibinfo {year} {2021})}\BibitemShut {NoStop}%
\bibitem [{\citenamefont {McEwen}\ \emph {et~al.}(2021)\citenamefont {McEwen}, \citenamefont {Kafri}, \citenamefont {Chen}, \citenamefont {Atalaya}, \citenamefont {Satzinger}, \citenamefont {Quintana}, \citenamefont {Klimov}, \citenamefont {Sank}, \citenamefont {Gidney}, \citenamefont {Fowler}, \citenamefont {Arute}, \citenamefont {Arya}, \citenamefont {Buckley}, \citenamefont {Burkett}, \citenamefont {Bushnell}, \citenamefont {Chiaro}, \citenamefont {Collins}, \citenamefont {Demura}, \citenamefont {Dunsworth}, \citenamefont {Erickson}, \citenamefont {Foxen}, \citenamefont {Giustina}, \citenamefont {Huang}, \citenamefont {Hong}, \citenamefont {Jeffrey}, \citenamefont {Kim}, \citenamefont {Kechedzhi}, \citenamefont {Kostritsa}, \citenamefont {Laptev}, \citenamefont {Megrant}, \citenamefont {Mi}, \citenamefont {Mutus}, \citenamefont {Naaman}, \citenamefont {Neeley}, \citenamefont {Neill}, \citenamefont {Niu}, \citenamefont {Paler}, \citenamefont {Redd}, \citenamefont {Roushan}, \citenamefont {White},
  \citenamefont {Yao}, \citenamefont {Yeh}, \citenamefont {Zalcman}, \citenamefont {Chen}, \citenamefont {Smelyanskiy}, \citenamefont {Martinis}, \citenamefont {Neven}, \citenamefont {Kelly}, \citenamefont {Korotkov}, \citenamefont {Petukhov},\ and\ \citenamefont {Barends}}]{McEwen2021}%
  \BibitemOpen
  \bibfield  {author} {\bibinfo {author} {\bibfnamefont {M.}~\bibnamefont {McEwen}}, \bibinfo {author} {\bibfnamefont {D.}~\bibnamefont {Kafri}}, \bibinfo {author} {\bibfnamefont {J.}~\bibnamefont {Chen}}, \bibinfo {author} {\bibfnamefont {J.}~\bibnamefont {Atalaya}}, \bibinfo {author} {\bibfnamefont {K.}~\bibnamefont {Satzinger}}, \bibinfo {author} {\bibfnamefont {C.}~\bibnamefont {Quintana}}, \bibinfo {author} {\bibfnamefont {P.~V.}\ \bibnamefont {Klimov}}, \bibinfo {author} {\bibfnamefont {D.}~\bibnamefont {Sank}}, \bibinfo {author} {\bibfnamefont {C.~M.}\ \bibnamefont {Gidney}}, \bibinfo {author} {\bibfnamefont {A.}~\bibnamefont {Fowler}}, \bibinfo {author} {\bibfnamefont {F.~C.}\ \bibnamefont {Arute}}, \bibinfo {author} {\bibfnamefont {K.}~\bibnamefont {Arya}}, \bibinfo {author} {\bibfnamefont {B.~B.}\ \bibnamefont {Buckley}}, \bibinfo {author} {\bibfnamefont {B.}~\bibnamefont {Burkett}}, \bibinfo {author} {\bibfnamefont {N.}~\bibnamefont {Bushnell}}, \bibinfo {author} {\bibfnamefont {B.}~\bibnamefont
  {Chiaro}}, \bibinfo {author} {\bibfnamefont {R.}~\bibnamefont {Collins}}, \bibinfo {author} {\bibfnamefont {S.}~\bibnamefont {Demura}}, \bibinfo {author} {\bibfnamefont {A.}~\bibnamefont {Dunsworth}}, \bibinfo {author} {\bibfnamefont {C.}~\bibnamefont {Erickson}}, \bibinfo {author} {\bibfnamefont {B.~R.}\ \bibnamefont {Foxen}}, \bibinfo {author} {\bibfnamefont {M.}~\bibnamefont {Giustina}}, \bibinfo {author} {\bibfnamefont {T.}~\bibnamefont {Huang}}, \bibinfo {author} {\bibfnamefont {S.}~\bibnamefont {Hong}}, \bibinfo {author} {\bibfnamefont {E.}~\bibnamefont {Jeffrey}}, \bibinfo {author} {\bibfnamefont {S.}~\bibnamefont {Kim}}, \bibinfo {author} {\bibfnamefont {K.}~\bibnamefont {Kechedzhi}}, \bibinfo {author} {\bibfnamefont {F.}~\bibnamefont {Kostritsa}}, \bibinfo {author} {\bibfnamefont {P.}~\bibnamefont {Laptev}}, \bibinfo {author} {\bibfnamefont {A.}~\bibnamefont {Megrant}}, \bibinfo {author} {\bibfnamefont {X.}~\bibnamefont {Mi}}, \bibinfo {author} {\bibfnamefont {J.}~\bibnamefont {Mutus}}, \bibinfo
  {author} {\bibfnamefont {O.}~\bibnamefont {Naaman}}, \bibinfo {author} {\bibfnamefont {M.}~\bibnamefont {Neeley}}, \bibinfo {author} {\bibfnamefont {C.}~\bibnamefont {Neill}}, \bibinfo {author} {\bibfnamefont {M.~Y.}\ \bibnamefont {Niu}}, \bibinfo {author} {\bibfnamefont {A.}~\bibnamefont {Paler}}, \bibinfo {author} {\bibfnamefont {N.}~\bibnamefont {Redd}}, \bibinfo {author} {\bibfnamefont {P.}~\bibnamefont {Roushan}}, \bibinfo {author} {\bibfnamefont {T.}~\bibnamefont {White}}, \bibinfo {author} {\bibfnamefont {J.}~\bibnamefont {Yao}}, \bibinfo {author} {\bibfnamefont {P.}~\bibnamefont {Yeh}}, \bibinfo {author} {\bibfnamefont {A.~J.}\ \bibnamefont {Zalcman}}, \bibinfo {author} {\bibfnamefont {Y.}~\bibnamefont {Chen}}, \bibinfo {author} {\bibfnamefont {V.}~\bibnamefont {Smelyanskiy}}, \bibinfo {author} {\bibfnamefont {J.}~\bibnamefont {Martinis}}, \bibinfo {author} {\bibfnamefont {H.}~\bibnamefont {Neven}}, \bibinfo {author} {\bibfnamefont {J.}~\bibnamefont {Kelly}}, \bibinfo {author} {\bibfnamefont
  {A.}~\bibnamefont {Korotkov}}, \bibinfo {author} {\bibfnamefont {A.~G.}\ \bibnamefont {Petukhov}},\ and\ \bibinfo {author} {\bibfnamefont {R.}~\bibnamefont {Barends}},\ }\bibfield  {title} {\bibinfo {title} {Removing leakage-induced correlated errors in superconducting quantum error correction},\ }\href {https://doi.org/10.1038/s41467-021-21982-y} {\bibfield  {journal} {\bibinfo  {journal} {Nature Commun.}\ }\textbf {\bibinfo {volume} {12}},\ \bibinfo {pages} {1761} (\bibinfo {year} {2021})}\BibitemShut {NoStop}%
\bibitem [{\citenamefont {Zou}\ \emph {et~al.}(2024)\citenamefont {Zou}, \citenamefont {Bosco},\ and\ \citenamefont {Loss}}]{Zou2024}%
  \BibitemOpen
  \bibfield  {author} {\bibinfo {author} {\bibfnamefont {J.}~\bibnamefont {Zou}}, \bibinfo {author} {\bibfnamefont {S.}~\bibnamefont {Bosco}},\ and\ \bibinfo {author} {\bibfnamefont {D.}~\bibnamefont {Loss}},\ }\bibfield  {title} {\bibinfo {title} {{Spatially correlated classical and quantum noise in driven qubits}},\ }\href {https://doi.org/10.1038/s41534-024-00842-9} {\bibfield  {journal} {\bibinfo  {journal} {npj Quantum Inf.}\ }\textbf {\bibinfo {volume} {10}},\ \bibinfo {pages} {46} (\bibinfo {year} {2024})}\BibitemShut {NoStop}%
\bibitem [{\citenamefont {Postema}\ and\ \citenamefont {Kokkelmans}(2024)}]{Postema2024}%
  \BibitemOpen
  \bibfield  {author} {\bibinfo {author} {\bibfnamefont {J.~J.}\ \bibnamefont {Postema}}\ and\ \bibinfo {author} {\bibfnamefont {S.}~\bibnamefont {Kokkelmans}},\ }\bibfield  {title} {\bibinfo {title} {Geometrical approach to logical qubit fidelities of neutral atom {CSS} codes},\ }\href {https://arxiv.org/abs/2409.04324} {\bibfield  {journal} {\bibinfo  {journal} {arXiv:2409.04324}\ } (\bibinfo {year} {2024})}\BibitemShut {NoStop}%
\bibitem [{\citenamefont {Hoyos}\ \emph {et~al.}(2011)\citenamefont {Hoyos}, \citenamefont {Laflorencie}, \citenamefont {Vieira},\ and\ \citenamefont {Vojta}}]{Hoyos2011}%
  \BibitemOpen
  \bibfield  {author} {\bibinfo {author} {\bibfnamefont {J.~A.}\ \bibnamefont {Hoyos}}, \bibinfo {author} {\bibfnamefont {N.}~\bibnamefont {Laflorencie}}, \bibinfo {author} {\bibfnamefont {A.~P.}\ \bibnamefont {Vieira}},\ and\ \bibinfo {author} {\bibfnamefont {T.}~\bibnamefont {Vojta}},\ }\bibfield  {title} {\bibinfo {title} {Protecting clean critical points by local disorder correlations},\ }\href {https://doi.org/10.1209/0295-5075/93/30004} {\bibfield  {journal} {\bibinfo  {journal} {Europhys. Lett.}\ }\textbf {\bibinfo {volume} {93}},\ \bibinfo {pages} {30004} (\bibinfo {year} {2011})}\BibitemShut {NoStop}%
\bibitem [{\citenamefont {Bonzom}\ \emph {et~al.}(2013)\citenamefont {Bonzom}, \citenamefont {Gurau},\ and\ \citenamefont {Smerlak}}]{Bonzom2013}%
  \BibitemOpen
  \bibfield  {author} {\bibinfo {author} {\bibfnamefont {V.}~\bibnamefont {Bonzom}}, \bibinfo {author} {\bibfnamefont {R.}~\bibnamefont {Gurau}},\ and\ \bibinfo {author} {\bibfnamefont {M.}~\bibnamefont {Smerlak}},\ }\bibfield  {title} {\bibinfo {title} {Universality in $p$-spin glasses with correlated disorder},\ }\href {https://doi.org/10.1088/1742-5468/2013/02/L02003} {\bibfield  {journal} {\bibinfo  {journal} {J. Stat. Mech.}\ }\textbf {\bibinfo {volume} {2013}},\ \bibinfo {pages} {L02003} (\bibinfo {year} {2013})}\BibitemShut {NoStop}%
\bibitem [{\citenamefont {Cavaliere}\ and\ \citenamefont {Pelissetto}(2019)}]{Cavaliere_2019}%
  \BibitemOpen
  \bibfield  {author} {\bibinfo {author} {\bibfnamefont {A.~G.}\ \bibnamefont {Cavaliere}}\ and\ \bibinfo {author} {\bibfnamefont {A.}~\bibnamefont {Pelissetto}},\ }\bibfield  {title} {\bibinfo {title} {Disordered {Ising} model with correlated frustration},\ }\href {https://doi.org/10.1088/1751-8121/ab10f9} {\bibfield  {journal} {\bibinfo  {journal} {J. Phys. A}\ }\textbf {\bibinfo {volume} {52}},\ \bibinfo {pages} {174002} (\bibinfo {year} {2019})}\BibitemShut {NoStop}%
\bibitem [{\citenamefont {M\"unster}\ \emph {et~al.}(2021)\citenamefont {M\"unster}, \citenamefont {Norrenbrock}, \citenamefont {Hartmann},\ and\ \citenamefont {Young}}]{Munster2021}%
  \BibitemOpen
  \bibfield  {author} {\bibinfo {author} {\bibfnamefont {L.}~\bibnamefont {M\"unster}}, \bibinfo {author} {\bibfnamefont {C.}~\bibnamefont {Norrenbrock}}, \bibinfo {author} {\bibfnamefont {A.~K.}\ \bibnamefont {Hartmann}},\ and\ \bibinfo {author} {\bibfnamefont {A.~P.}\ \bibnamefont {Young}},\ }\bibfield  {title} {\bibinfo {title} {Ordering behavior of the two-dimensional {Ising} spin glass with long-range correlated disorder},\ }\href {https://doi.org/10.1103/PhysRevE.103.042117} {\bibfield  {journal} {\bibinfo  {journal} {Phys. Rev. E}\ }\textbf {\bibinfo {volume} {103}},\ \bibinfo {pages} {042117} (\bibinfo {year} {2021})}\BibitemShut {NoStop}%
\bibitem [{\citenamefont {Nishimori}(2022)}]{Nishimori2022}%
  \BibitemOpen
  \bibfield  {author} {\bibinfo {author} {\bibfnamefont {H.}~\bibnamefont {Nishimori}},\ }\bibfield  {title} {\bibinfo {title} {Analyticity of the energy in an {Ising} spin glass with correlated disorder},\ }\href {https://doi.org/10.1088/1751-8121/ac44ef} {\bibfield  {journal} {\bibinfo  {journal} {J. Phys. A}\ }\textbf {\bibinfo {volume} {55}},\ \bibinfo {pages} {045001} (\bibinfo {year} {2022})}\BibitemShut {NoStop}%
\bibitem [{\citenamefont {Alberici}\ \emph {et~al.}(2021)\citenamefont {Alberici}, \citenamefont {Camilli}, \citenamefont {Contucci},\ and\ \citenamefont {Mingione}}]{alberici2021multi}%
  \BibitemOpen
  \bibfield  {author} {\bibinfo {author} {\bibfnamefont {D.}~\bibnamefont {Alberici}}, \bibinfo {author} {\bibfnamefont {F.}~\bibnamefont {Camilli}}, \bibinfo {author} {\bibfnamefont {P.}~\bibnamefont {Contucci}},\ and\ \bibinfo {author} {\bibfnamefont {E.}~\bibnamefont {Mingione}},\ }\bibfield  {title} {\bibinfo {title} {The multi-species mean-field spin-glass on the {Nishimori} line},\ }\href {https://doi.org/10.1007/s10955-020-02684-z} {\bibfield  {journal} {\bibinfo  {journal} {J. Stat. Phys.}\ }\textbf {\bibinfo {volume} {182}},\ \bibinfo {pages} {2} (\bibinfo {year} {2021})}\BibitemShut {NoStop}%
\bibitem [{\citenamefont {Tanaka}(2022)}]{tanaka2022nishimori}%
  \BibitemOpen
  \bibfield  {author} {\bibinfo {author} {\bibfnamefont {K.}~\bibnamefont {Tanaka}},\ }\bibfield  {title} {\bibinfo {title} {Nishimori line for general spin random-bond {Ising} systems on honeycomb lattice},\ }\href {https://doi.org/10.7566/JPSJ.91.115003} {\bibfield  {journal} {\bibinfo  {journal} {J. Phys. Soc. Jpn.}\ }\textbf {\bibinfo {volume} {91}},\ \bibinfo {pages} {115003} (\bibinfo {year} {2022})}\BibitemShut {NoStop}%
\bibitem [{\citenamefont {Garban}\ and\ \citenamefont {Spencer}(2022)}]{garban2022continuous}%
  \BibitemOpen
  \bibfield  {author} {\bibinfo {author} {\bibfnamefont {C.}~\bibnamefont {Garban}}\ and\ \bibinfo {author} {\bibfnamefont {T.}~\bibnamefont {Spencer}},\ }\bibfield  {title} {\bibinfo {title} {Continuous symmetry breaking along the {Nishimori} line},\ }\href {https://doi.org/10.1063/5.0087024} {\bibfield  {journal} {\bibinfo  {journal} {J. Math. Phys.}\ }\textbf {\bibinfo {volume} {63}},\ \bibinfo {pages} {093302} (\bibinfo {year} {2022})}\BibitemShut {NoStop}%
\bibitem [{\citenamefont {Okuyama}\ and\ \citenamefont {Ohzeki}(2023{\natexlab{a}})}]{okuyama2023}%
  \BibitemOpen
  \bibfield  {author} {\bibinfo {author} {\bibfnamefont {M.}~\bibnamefont {Okuyama}}\ and\ \bibinfo {author} {\bibfnamefont {M.}~\bibnamefont {Ohzeki}},\ }\bibfield  {title} {\bibinfo {title} {{Gibbs--Bogoliubov} inequality on the {Nishimori} line},\ }\href {https://doi.org/10.7566/JPSJ.92.084002} {\bibfield  {journal} {\bibinfo  {journal} {J. Phys. Soc. Jpn.}\ }\textbf {\bibinfo {volume} {92}},\ \bibinfo {pages} {084002} (\bibinfo {year} {2023}{\natexlab{a}})}\BibitemShut {NoStop}%
\bibitem [{\citenamefont {Okuyama}\ and\ \citenamefont {Ohzeki}(2023{\natexlab{b}})}]{okuyama2023mean}%
  \BibitemOpen
  \bibfield  {author} {\bibinfo {author} {\bibfnamefont {M.}~\bibnamefont {Okuyama}}\ and\ \bibinfo {author} {\bibfnamefont {M.}~\bibnamefont {Ohzeki}},\ }\bibfield  {title} {\bibinfo {title} {Mean-field theory is exact for {Ising} spin glass models with {Kac} potential in non-additive limit on {Nishimori} line},\ }\href {https://doi.org/10.1088/1751-8121/ace6e4} {\bibfield  {journal} {\bibinfo  {journal} {J. Phys. A}\ }\textbf {\bibinfo {volume} {56}},\ \bibinfo {pages} {325003} (\bibinfo {year} {2023}{\natexlab{b}})}\BibitemShut {NoStop}%
\bibitem [{\citenamefont {Itoi}\ and\ \citenamefont {Sakamoto}(2023)}]{itoi2023boundedness}%
  \BibitemOpen
  \bibfield  {author} {\bibinfo {author} {\bibfnamefont {C.}~\bibnamefont {Itoi}}\ and\ \bibinfo {author} {\bibfnamefont {Y.}~\bibnamefont {Sakamoto}},\ }\bibfield  {title} {\bibinfo {title} {Boundedness of susceptibility in spin glass transition of transverse field mixed $p$-spin glass models},\ }\href {https://doi.org/10.7566/JPSJ.92.064004} {\bibfield  {journal} {\bibinfo  {journal} {J. Phys. Soc. Jpn.}\ }\textbf {\bibinfo {volume} {92}},\ \bibinfo {pages} {064004} (\bibinfo {year} {2023})}\BibitemShut {NoStop}%
\bibitem [{\citenamefont {Placke}\ and\ \citenamefont {Breuckmann}(2023)}]{Placke2023}%
  \BibitemOpen
  \bibfield  {author} {\bibinfo {author} {\bibfnamefont {B.}~\bibnamefont {Placke}}\ and\ \bibinfo {author} {\bibfnamefont {N.~P.}\ \bibnamefont {Breuckmann}},\ }\bibfield  {title} {\bibinfo {title} {Random-bond {Ising} model and its dual in hyperbolic spaces},\ }\href {https://doi.org/10.1103/PhysRevE.107.024125} {\bibfield  {journal} {\bibinfo  {journal} {Phys. Rev. E}\ }\textbf {\bibinfo {volume} {107}},\ \bibinfo {pages} {024125} (\bibinfo {year} {2023})}\BibitemShut {NoStop}%
\bibitem [{\citenamefont {Terasawa}\ and\ \citenamefont {Ozeki}(2023)}]{terasawa2023dynamical}%
  \BibitemOpen
  \bibfield  {author} {\bibinfo {author} {\bibfnamefont {Y.}~\bibnamefont {Terasawa}}\ and\ \bibinfo {author} {\bibfnamefont {Y.}~\bibnamefont {Ozeki}},\ }\bibfield  {title} {\bibinfo {title} {Dynamical scaling analysis for {$\pm J$} {Ising} model in three dimensions},\ }\href {https://doi.org/10.7566/JPSJ.92.074003} {\bibfield  {journal} {\bibinfo  {journal} {J. Phys. Soc. Jpn.}\ }\textbf {\bibinfo {volume} {92}},\ \bibinfo {pages} {074003} (\bibinfo {year} {2023})}\BibitemShut {NoStop}%
\bibitem [{\citenamefont {M\"unster}\ and\ \citenamefont {Weigel}(2023)}]{Munster2023}%
  \BibitemOpen
  \bibfield  {author} {\bibinfo {author} {\bibfnamefont {L.}~\bibnamefont {M\"unster}}\ and\ \bibinfo {author} {\bibfnamefont {M.}~\bibnamefont {Weigel}},\ }\bibfield  {title} {\bibinfo {title} {Cluster percolation in the two-dimensional {Ising} spin glass},\ }\href {https://doi.org/10.1103/PhysRevE.107.054103} {\bibfield  {journal} {\bibinfo  {journal} {Phys. Rev. E}\ }\textbf {\bibinfo {volume} {107}},\ \bibinfo {pages} {054103} (\bibinfo {year} {2023})}\BibitemShut {NoStop}%
\bibitem [{\citenamefont {Camilli}\ \emph {et~al.}(2023)\citenamefont {Camilli}, \citenamefont {Contucci},\ and\ \citenamefont {Mingione}}]{camilli2023onset}%
  \BibitemOpen
  \bibfield  {author} {\bibinfo {author} {\bibfnamefont {F.}~\bibnamefont {Camilli}}, \bibinfo {author} {\bibfnamefont {P.}~\bibnamefont {Contucci}},\ and\ \bibinfo {author} {\bibfnamefont {E.}~\bibnamefont {Mingione}},\ }\bibfield  {title} {\bibinfo {title} {The onset of {Parisi's} complexity in a mismatched inference problem},\ }\href {https://doi.org/10.3390/e26010042} {\bibfield  {journal} {\bibinfo  {journal} {Entropy}\ }\textbf {\bibinfo {volume} {26}},\ \bibinfo {pages} {42} (\bibinfo {year} {2023})}\BibitemShut {NoStop}%
\bibitem [{\citenamefont {Itoi}\ and\ \citenamefont {Sakamoto}(2024)}]{itoi2024gauge}%
  \BibitemOpen
  \bibfield  {author} {\bibinfo {author} {\bibfnamefont {C.}~\bibnamefont {Itoi}}\ and\ \bibinfo {author} {\bibfnamefont {Y.}~\bibnamefont {Sakamoto}},\ }\bibfield  {title} {\bibinfo {title} {Gauge theory for quantum {XYZ} spin glasses},\ }\href {https://doi.org/10.1088/1751-8121/ad1a1d} {\bibfield  {journal} {\bibinfo  {journal} {J. Phys. A}\ }\textbf {\bibinfo {volume} {57}},\ \bibinfo {pages} {045001} (\bibinfo {year} {2024})}\BibitemShut {NoStop}%
\bibitem [{\citenamefont {Sirenko}\ \emph {et~al.}(2024)\citenamefont {Sirenko}, \citenamefont {Bartolom{\'e}~Usieto},\ and\ \citenamefont {Bartolom{\'e}}}]{sirenko2024paradigm}%
  \BibitemOpen
  \bibfield  {author} {\bibinfo {author} {\bibfnamefont {V.}~\bibnamefont {Sirenko}}, \bibinfo {author} {\bibfnamefont {F.}~\bibnamefont {Bartolom{\'e}~Usieto}},\ and\ \bibinfo {author} {\bibfnamefont {J.}~\bibnamefont {Bartolom{\'e}}},\ }\bibfield  {title} {\bibinfo {title} {The paradigm of magnetic molecule in quantum matter: {Slow} molecular spin relaxation},\ }\href {https://doi.org/10.1063/10.0026056} {\bibfield  {journal} {\bibinfo  {journal} {Low Temp. Phys.}\ }\textbf {\bibinfo {volume} {50}},\ \bibinfo {pages} {431} (\bibinfo {year} {2024})}\BibitemShut {NoStop}%
\bibitem [{\citenamefont {Agrawal}\ \emph {et~al.}(2023)\citenamefont {Agrawal}, \citenamefont {Cugliandolo}, \citenamefont {Faoro}, \citenamefont {Ioffe},\ and\ \citenamefont {Picco}}]{Agrawal2023}%
  \BibitemOpen
  \bibfield  {author} {\bibinfo {author} {\bibfnamefont {R.}~\bibnamefont {Agrawal}}, \bibinfo {author} {\bibfnamefont {L.~F.}\ \bibnamefont {Cugliandolo}}, \bibinfo {author} {\bibfnamefont {L.}~\bibnamefont {Faoro}}, \bibinfo {author} {\bibfnamefont {L.~B.}\ \bibnamefont {Ioffe}},\ and\ \bibinfo {author} {\bibfnamefont {M.}~\bibnamefont {Picco}},\ }\bibfield  {title} {\bibinfo {title} {Nonequilibrium critical dynamics of the two-dimensional {$\ifmmode\pm\else\textpm\fi{}J$ Ising} model},\ }\href {https://doi.org/10.1103/PhysRevE.108.064131} {\bibfield  {journal} {\bibinfo  {journal} {Phys. Rev. E}\ }\textbf {\bibinfo {volume} {108}},\ \bibinfo {pages} {064131} (\bibinfo {year} {2023})}\BibitemShut {NoStop}%
\bibitem [{\citenamefont {Braunstein}\ \emph {et~al.}(2023)\citenamefont {Braunstein}, \citenamefont {Budzynski},\ and\ \citenamefont {Mariani}}]{Braunstein2023}%
  \BibitemOpen
  \bibfield  {author} {\bibinfo {author} {\bibfnamefont {A.}~\bibnamefont {Braunstein}}, \bibinfo {author} {\bibfnamefont {L.}~\bibnamefont {Budzynski}},\ and\ \bibinfo {author} {\bibfnamefont {M.}~\bibnamefont {Mariani}},\ }\bibfield  {title} {\bibinfo {title} {Statistical mechanics of inference in epidemic spreading},\ }\href {https://doi.org/10.1103/PhysRevE.108.064302} {\bibfield  {journal} {\bibinfo  {journal} {Phys. Rev. E}\ }\textbf {\bibinfo {volume} {108}},\ \bibinfo {pages} {064302} (\bibinfo {year} {2023})}\BibitemShut {NoStop}%
\bibitem [{\citenamefont {Crotti}\ and\ \citenamefont {Braunstein}(2023)}]{crotti2023matrix}%
  \BibitemOpen
  \bibfield  {author} {\bibinfo {author} {\bibfnamefont {S.}~\bibnamefont {Crotti}}\ and\ \bibinfo {author} {\bibfnamefont {A.}~\bibnamefont {Braunstein}},\ }\bibfield  {title} {\bibinfo {title} {Matrix product belief propagation for reweighted stochastic dynamics over graphs},\ }\href {https://doi.org/10.1073/pnas.2307935120} {\bibfield  {journal} {\bibinfo  {journal} {Proc. Nat. Acad. Sci.}\ }\textbf {\bibinfo {volume} {120}},\ \bibinfo {pages} {e2307935120} (\bibinfo {year} {2023})}\BibitemShut {NoStop}%
\bibitem [{\citenamefont {Ghio}\ \emph {et~al.}(2024)\citenamefont {Ghio}, \citenamefont {Dandi}, \citenamefont {Krzakala},\ and\ \citenamefont {Zdeborov{\'a}}}]{ghio2024sampling}%
  \BibitemOpen
  \bibfield  {author} {\bibinfo {author} {\bibfnamefont {D.}~\bibnamefont {Ghio}}, \bibinfo {author} {\bibfnamefont {Y.}~\bibnamefont {Dandi}}, \bibinfo {author} {\bibfnamefont {F.}~\bibnamefont {Krzakala}},\ and\ \bibinfo {author} {\bibfnamefont {L.}~\bibnamefont {Zdeborov{\'a}}},\ }\bibfield  {title} {\bibinfo {title} {Sampling with flows, diffusion, and autoregressive neural networks from a spin-glass perspective},\ }\href {https://doi.org/10.1073/pnas.2311810121} {\bibfield  {journal} {\bibinfo  {journal} {Proc. Nat. Acad. Sci.}\ }\textbf {\bibinfo {volume} {121}},\ \bibinfo {pages} {e2311810121} (\bibinfo {year} {2024})}\BibitemShut {NoStop}%
\bibitem [{\citenamefont {Genovese}(2023)}]{genovese2023minimax}%
  \BibitemOpen
  \bibfield  {author} {\bibinfo {author} {\bibfnamefont {G.}~\bibnamefont {Genovese}},\ }\bibfield  {title} {\bibinfo {title} {Minimax formula for the replica symmetric free energy of deep restricted {Boltzmann} machines},\ }\href {https://doi.org/10.1214/22-AAP1868} {\bibfield  {journal} {\bibinfo  {journal} {The Annals of Applied Probability}\ }\textbf {\bibinfo {volume} {33}},\ \bibinfo {pages} {2324} (\bibinfo {year} {2023})}\BibitemShut {NoStop}%
\bibitem [{\citenamefont {Ettori}\ \emph {et~al.}(2023)\citenamefont {Ettori}, \citenamefont {Perani}, \citenamefont {Turzi},\ and\ \citenamefont {Biscari}}]{ettori2023finite}%
  \BibitemOpen
  \bibfield  {author} {\bibinfo {author} {\bibfnamefont {F.}~\bibnamefont {Ettori}}, \bibinfo {author} {\bibfnamefont {F.}~\bibnamefont {Perani}}, \bibinfo {author} {\bibfnamefont {S.}~\bibnamefont {Turzi}},\ and\ \bibinfo {author} {\bibfnamefont {P.}~\bibnamefont {Biscari}},\ }\bibfield  {title} {\bibinfo {title} {Finite-temperature avalanches in {2D} disordered {Ising} models},\ }\href {https://doi.org/10.1007/s10955-023-03098-3} {\bibfield  {journal} {\bibinfo  {journal} {J. Stat. Phys.}\ }\textbf {\bibinfo {volume} {190}},\ \bibinfo {pages} {89} (\bibinfo {year} {2023})}\BibitemShut {NoStop}%
\bibitem [{\citenamefont {Angelini}\ and\ \citenamefont {Ricci-Tersenghi}(2023)}]{Angelini2023}%
  \BibitemOpen
  \bibfield  {author} {\bibinfo {author} {\bibfnamefont {M.~C.}\ \bibnamefont {Angelini}}\ and\ \bibinfo {author} {\bibfnamefont {F.}~\bibnamefont {Ricci-Tersenghi}},\ }\bibfield  {title} {\bibinfo {title} {Limits and performances of algorithms based on simulated annealing in solving sparse hard inference problems},\ }\href {https://doi.org/10.1103/PhysRevX.13.021011} {\bibfield  {journal} {\bibinfo  {journal} {Phys. Rev. X}\ }\textbf {\bibinfo {volume} {13}},\ \bibinfo {pages} {021011} (\bibinfo {year} {2023})}\BibitemShut {NoStop}%
\bibitem [{\citenamefont {Zdeborov{\'a}}\ and\ \citenamefont {Krzakala}(2016)}]{Zdeborova2016}%
  \BibitemOpen
  \bibfield  {author} {\bibinfo {author} {\bibfnamefont {L.}~\bibnamefont {Zdeborov{\'a}}}\ and\ \bibinfo {author} {\bibfnamefont {F.}~\bibnamefont {Krzakala}},\ }\bibfield  {title} {\bibinfo {title} {Statistical physics of inference: Thresholds and algorithms},\ }\href {https://doi.org/10.1080/00018732.2016.1211393} {\bibfield  {journal} {\bibinfo  {journal} {Adv. Phys.}\ }\textbf {\bibinfo {volume} {65}},\ \bibinfo {pages} {453} (\bibinfo {year} {2016})}\BibitemShut {NoStop}%
\bibitem [{\citenamefont {Zhu}\ \emph {et~al.}(2023)\citenamefont {Zhu}, \citenamefont {Tantivasadakarn}, \citenamefont {Vishwanath}, \citenamefont {Trebst},\ and\ \citenamefont {Verresen}}]{Zhu2023}%
  \BibitemOpen
  \bibfield  {author} {\bibinfo {author} {\bibfnamefont {G.-Y.}\ \bibnamefont {Zhu}}, \bibinfo {author} {\bibfnamefont {N.}~\bibnamefont {Tantivasadakarn}}, \bibinfo {author} {\bibfnamefont {A.}~\bibnamefont {Vishwanath}}, \bibinfo {author} {\bibfnamefont {S.}~\bibnamefont {Trebst}},\ and\ \bibinfo {author} {\bibfnamefont {R.}~\bibnamefont {Verresen}},\ }\bibfield  {title} {\bibinfo {title} {Nishimori's cat: Stable long-range entanglement from finite-depth unitaries and weak measurements},\ }\href {https://doi.org/10.1103/PhysRevLett.131.200201} {\bibfield  {journal} {\bibinfo  {journal} {Phys. Rev. Lett.}\ }\textbf {\bibinfo {volume} {131}},\ \bibinfo {pages} {200201} (\bibinfo {year} {2023})}\BibitemShut {NoStop}%
\bibitem [{\citenamefont {Chen}\ \emph {et~al.}(2023)\citenamefont {Chen}, \citenamefont {Zhu}, \citenamefont {Verresen}, \citenamefont {Seif}, \citenamefont {B\"aumer}, \citenamefont {Layden}, \citenamefont {Tantivasadakarn}, \citenamefont {Zhu}, \citenamefont {Sheldon}, \citenamefont {Vishwanath}, \citenamefont {Trebst},\ and\ \citenamefont {Kandala}}]{chen2023}%
  \BibitemOpen
  \bibfield  {author} {\bibinfo {author} {\bibfnamefont {E.~H.}\ \bibnamefont {Chen}}, \bibinfo {author} {\bibfnamefont {G.-Y.}\ \bibnamefont {Zhu}}, \bibinfo {author} {\bibfnamefont {R.}~\bibnamefont {Verresen}}, \bibinfo {author} {\bibfnamefont {A.}~\bibnamefont {Seif}}, \bibinfo {author} {\bibfnamefont {E.}~\bibnamefont {B\"aumer}}, \bibinfo {author} {\bibfnamefont {D.}~\bibnamefont {Layden}}, \bibinfo {author} {\bibfnamefont {N.}~\bibnamefont {Tantivasadakarn}}, \bibinfo {author} {\bibfnamefont {G.}~\bibnamefont {Zhu}}, \bibinfo {author} {\bibfnamefont {S.}~\bibnamefont {Sheldon}}, \bibinfo {author} {\bibfnamefont {A.}~\bibnamefont {Vishwanath}}, \bibinfo {author} {\bibfnamefont {S.}~\bibnamefont {Trebst}},\ and\ \bibinfo {author} {\bibfnamefont {A.}~\bibnamefont {Kandala}},\ }\bibfield  {title} {\bibinfo {title} {Realizing the {Nishimori} transition across the error threshold for constant-depth quantum circuits},\ }\href {https://arxiv.org/abs/2309.02863} {\bibfield  {journal} {\bibinfo  {journal}
  {arxiv:2309.02863}\ } (\bibinfo {year} {2023})}\BibitemShut {NoStop}%
\bibitem [{\citenamefont {Lee}\ \emph {et~al.}(2022)\citenamefont {Lee}, \citenamefont {Ji}, \citenamefont {Bi},\ and\ \citenamefont {Fisher}}]{lee2022}%
  \BibitemOpen
  \bibfield  {author} {\bibinfo {author} {\bibfnamefont {J.~Y.}\ \bibnamefont {Lee}}, \bibinfo {author} {\bibfnamefont {W.}~\bibnamefont {Ji}}, \bibinfo {author} {\bibfnamefont {Z.}~\bibnamefont {Bi}},\ and\ \bibinfo {author} {\bibfnamefont {M.~P.~A.}\ \bibnamefont {Fisher}},\ }\bibfield  {title} {\bibinfo {title} {Decoding measurement-prepared quantum phases and transitions: from {Ising} model to gauge theory, and beyond},\ }\href {https://arxiv.org/abs/2208.11699} {\bibfield  {journal} {\bibinfo  {journal} {arxiv:2208.11699}\ } (\bibinfo {year} {2022})}\BibitemShut {NoStop}%
\bibitem [{\citenamefont {Nishimori}\ and\ \citenamefont {Sherrington}(2001)}]{Nishimori2001b}%
  \BibitemOpen
  \bibfield  {author} {\bibinfo {author} {\bibfnamefont {H.}~\bibnamefont {Nishimori}}\ and\ \bibinfo {author} {\bibfnamefont {D.}~\bibnamefont {Sherrington}},\ }\bibfield  {title} {\bibinfo {title} {Absence of replica symmetry breaking in a region of the phase diagram of the {Ising} spin glass},\ }\href {https://doi.org/10.1063/1.1358165} {\bibfield  {journal} {\bibinfo  {journal} {AIP Conf. Proc.}\ }\textbf {\bibinfo {volume} {553}},\ \bibinfo {pages} {67} (\bibinfo {year} {2001})}\BibitemShut {NoStop}%
\bibitem [{\citenamefont {Parisi}\ and\ \citenamefont {Rizzo}(2010)}]{Parisi2010}%
  \BibitemOpen
  \bibfield  {author} {\bibinfo {author} {\bibfnamefont {G.}~\bibnamefont {Parisi}}\ and\ \bibinfo {author} {\bibfnamefont {T.}~\bibnamefont {Rizzo}},\ }\bibfield  {title} {\bibinfo {title} {{Chaos in temperature in diluted mean-field spin-glass}},\ }\href {https://doi.org/10.1088/1751-8113/43/23/235003} {\bibfield  {journal} {\bibinfo  {journal} {J. Phys. A}\ }\textbf {\bibinfo {volume} {43}},\ \bibinfo {pages} {235003} (\bibinfo {year} {2010})}\BibitemShut {NoStop}%
\bibitem [{\citenamefont {Bray}\ and\ \citenamefont {Moore}(1987)}]{Bray1987}%
  \BibitemOpen
  \bibfield  {author} {\bibinfo {author} {\bibfnamefont {A.~J.}\ \bibnamefont {Bray}}\ and\ \bibinfo {author} {\bibfnamefont {M.~A.}\ \bibnamefont {Moore}},\ }\bibfield  {title} {\bibinfo {title} {Chaotic nature of the spin-glass phase},\ }\href {https://doi.org/10.1103/PhysRevLett.58.57} {\bibfield  {journal} {\bibinfo  {journal} {Phys. Rev. Lett.}\ }\textbf {\bibinfo {volume} {58}},\ \bibinfo {pages} {57} (\bibinfo {year} {1987})}\BibitemShut {NoStop}%
\bibitem [{\citenamefont {Banavar}\ and\ \citenamefont {Bray}(1987)}]{Banavar1987}%
  \BibitemOpen
  \bibfield  {author} {\bibinfo {author} {\bibfnamefont {J.~R.}\ \bibnamefont {Banavar}}\ and\ \bibinfo {author} {\bibfnamefont {A.~J.}\ \bibnamefont {Bray}},\ }\bibfield  {title} {\bibinfo {title} {Chaos in spin glasses: A renormalization-group study},\ }\href {https://doi.org/10.1103/PhysRevB.35.8888} {\bibfield  {journal} {\bibinfo  {journal} {Phys. Rev. B}\ }\textbf {\bibinfo {volume} {35}},\ \bibinfo {pages} {8888} (\bibinfo {year} {1987})}\BibitemShut {NoStop}%
\bibitem [{\citenamefont {Fisher}\ and\ \citenamefont {Huse}(1986)}]{Fisher1986}%
  \BibitemOpen
  \bibfield  {author} {\bibinfo {author} {\bibfnamefont {D.~S.}\ \bibnamefont {Fisher}}\ and\ \bibinfo {author} {\bibfnamefont {D.~A.}\ \bibnamefont {Huse}},\ }\bibfield  {title} {\bibinfo {title} {Ordered phase of short-range {Ising} spin-glasses},\ }\href {https://doi.org/10.1103/PhysRevLett.56.1601} {\bibfield  {journal} {\bibinfo  {journal} {Phys. Rev. Lett.}\ }\textbf {\bibinfo {volume} {56}},\ \bibinfo {pages} {1601} (\bibinfo {year} {1986})}\BibitemShut {NoStop}%
\bibitem [{\citenamefont {Fisher}\ and\ \citenamefont {Huse}(1988)}]{Fisher1988}%
  \BibitemOpen
  \bibfield  {author} {\bibinfo {author} {\bibfnamefont {D.~S.}\ \bibnamefont {Fisher}}\ and\ \bibinfo {author} {\bibfnamefont {D.~A.}\ \bibnamefont {Huse}},\ }\bibfield  {title} {\bibinfo {title} {{Equilibrium behavior of the spin-glass ordered phase}},\ }\href {https://doi.org/10.1103/PhysRevB.38.386} {\bibfield  {journal} {\bibinfo  {journal} {Phys. Rev. B}\ }\textbf {\bibinfo {volume} {38}},\ \bibinfo {pages} {386} (\bibinfo {year} {1988})}\BibitemShut {NoStop}%
\bibitem [{\citenamefont {Kondor}(1989)}]{Kondor1989}%
  \BibitemOpen
  \bibfield  {author} {\bibinfo {author} {\bibfnamefont {I.}~\bibnamefont {Kondor}},\ }\bibfield  {title} {\bibinfo {title} {On chaos in spin glasses},\ }\href {https://doi.org/10.1088/0305-4470/22/5/005} {\bibfield  {journal} {\bibinfo  {journal} {J. Phys. A}\ }\textbf {\bibinfo {volume} {22}},\ \bibinfo {pages} {L163} (\bibinfo {year} {1989})}\BibitemShut {NoStop}%
\bibitem [{\citenamefont {Ney-Nifle}\ and\ \citenamefont {Young}(1997)}]{Ney-Nifle_1997}%
  \BibitemOpen
  \bibfield  {author} {\bibinfo {author} {\bibfnamefont {M.}~\bibnamefont {Ney-Nifle}}\ and\ \bibinfo {author} {\bibfnamefont {A.~P.}\ \bibnamefont {Young}},\ }\bibfield  {title} {\bibinfo {title} {Chaos in a two-dimensional ising spin glass},\ }\href {https://doi.org/10.1088/0305-4470/30/15/017} {\bibfield  {journal} {\bibinfo  {journal} {J. Phys. A}\ }\textbf {\bibinfo {volume} {30}},\ \bibinfo {pages} {5311} (\bibinfo {year} {1997})}\BibitemShut {NoStop}%
\bibitem [{\citenamefont {Ney-Nifle}(1998)}]{Ney-Nifle1998}%
  \BibitemOpen
  \bibfield  {author} {\bibinfo {author} {\bibfnamefont {M.}~\bibnamefont {Ney-Nifle}},\ }\bibfield  {title} {\bibinfo {title} {Chaos and universality in a four-dimensional spin glass},\ }\href {https://doi.org/10.1103/PhysRevB.57.492} {\bibfield  {journal} {\bibinfo  {journal} {Phys. Rev. B}\ }\textbf {\bibinfo {volume} {57}},\ \bibinfo {pages} {492} (\bibinfo {year} {1998})}\BibitemShut {NoStop}%
\bibitem [{\citenamefont {Mathieu}\ \emph {et~al.}(2001)\citenamefont {Mathieu}, \citenamefont {J\"onsson}, \citenamefont {Nordblad}, \citenamefont {Katori},\ and\ \citenamefont {Ito}}]{Mathieu2001}%
  \BibitemOpen
  \bibfield  {author} {\bibinfo {author} {\bibfnamefont {R.}~\bibnamefont {Mathieu}}, \bibinfo {author} {\bibfnamefont {P.~E.}\ \bibnamefont {J\"onsson}}, \bibinfo {author} {\bibfnamefont {P.}~\bibnamefont {Nordblad}}, \bibinfo {author} {\bibfnamefont {H.~A.}\ \bibnamefont {Katori}},\ and\ \bibinfo {author} {\bibfnamefont {A.}~\bibnamefont {Ito}},\ }\bibfield  {title} {\bibinfo {title} {Memory and chaos in an {Ising} spin glass},\ }\href {https://doi.org/10.1103/PhysRevB.65.012411} {\bibfield  {journal} {\bibinfo  {journal} {Phys. Rev. B}\ }\textbf {\bibinfo {volume} {65}},\ \bibinfo {pages} {012411} (\bibinfo {year} {2001})}\BibitemShut {NoStop}%
\bibitem [{\citenamefont {Bouchaud}\ \emph {et~al.}(2001)\citenamefont {Bouchaud}, \citenamefont {Dupuis}, \citenamefont {Hammann},\ and\ \citenamefont {Vincent}}]{Bouchaud2001}%
  \BibitemOpen
  \bibfield  {author} {\bibinfo {author} {\bibfnamefont {J.-P.}\ \bibnamefont {Bouchaud}}, \bibinfo {author} {\bibfnamefont {V.}~\bibnamefont {Dupuis}}, \bibinfo {author} {\bibfnamefont {J.}~\bibnamefont {Hammann}},\ and\ \bibinfo {author} {\bibfnamefont {E.}~\bibnamefont {Vincent}},\ }\bibfield  {title} {\bibinfo {title} {Separation of time and length scales in spin glasses: Temperature as a microscope},\ }\href {https://doi.org/10.1103/PhysRevB.65.024439} {\bibfield  {journal} {\bibinfo  {journal} {Phys. Rev. B}\ }\textbf {\bibinfo {volume} {65}},\ \bibinfo {pages} {024439} (\bibinfo {year} {2001})}\BibitemShut {NoStop}%
\bibitem [{\citenamefont {Aspelmeier}\ \emph {et~al.}(2002)\citenamefont {Aspelmeier}, \citenamefont {Bray},\ and\ \citenamefont {Moore}}]{Aspelmeier2002}%
  \BibitemOpen
  \bibfield  {author} {\bibinfo {author} {\bibfnamefont {T.}~\bibnamefont {Aspelmeier}}, \bibinfo {author} {\bibfnamefont {A.~J.}\ \bibnamefont {Bray}},\ and\ \bibinfo {author} {\bibfnamefont {M.~A.}\ \bibnamefont {Moore}},\ }\bibfield  {title} {\bibinfo {title} {Why temperature chaos in spin glasses is hard to observe},\ }\href {https://doi.org/10.1103/PhysRevLett.89.197202} {\bibfield  {journal} {\bibinfo  {journal} {Phys. Rev. Lett.}\ }\textbf {\bibinfo {volume} {89}},\ \bibinfo {pages} {197202} (\bibinfo {year} {2002})}\BibitemShut {NoStop}%
\bibitem [{\citenamefont {Rizzo}\ and\ \citenamefont {Crisanti}(2003)}]{Rizzo2003}%
  \BibitemOpen
  \bibfield  {author} {\bibinfo {author} {\bibfnamefont {T.}~\bibnamefont {Rizzo}}\ and\ \bibinfo {author} {\bibfnamefont {A.}~\bibnamefont {Crisanti}},\ }\bibfield  {title} {\bibinfo {title} {{Chaos in temperature in the Sherrington-Kirkpatrick model}},\ }\href {https://doi.org/10.1103/PhysRevLett.90.137201} {\bibfield  {journal} {\bibinfo  {journal} {Phys. Rev. Lett.}\ }\textbf {\bibinfo {volume} {90}},\ \bibinfo {pages} {4} (\bibinfo {year} {2003})}\BibitemShut {NoStop}%
\bibitem [{\citenamefont {Houdayer}\ and\ \citenamefont {Hartmann}(2004)}]{Houdayer2004}%
  \BibitemOpen
  \bibfield  {author} {\bibinfo {author} {\bibfnamefont {J.}~\bibnamefont {Houdayer}}\ and\ \bibinfo {author} {\bibfnamefont {A.~K.}\ \bibnamefont {Hartmann}},\ }\bibfield  {title} {\bibinfo {title} {Low-temperature behavior of two-dimensional gaussian {Ising} spin glasses},\ }\href {https://doi.org/10.1103/PhysRevB.70.014418} {\bibfield  {journal} {\bibinfo  {journal} {Phys. Rev. B}\ }\textbf {\bibinfo {volume} {70}},\ \bibinfo {pages} {014418} (\bibinfo {year} {2004})}\BibitemShut {NoStop}%
\bibitem [{\citenamefont {Katzgraber}\ and\ \citenamefont {Krza\ifmmode~\mbox{\c{}}\else \c{}\fi{}ka\l{}a}(2007)}]{Katzgraber2007b}%
  \BibitemOpen
  \bibfield  {author} {\bibinfo {author} {\bibfnamefont {H.~G.}\ \bibnamefont {Katzgraber}}\ and\ \bibinfo {author} {\bibfnamefont {F.}~\bibnamefont {Krza\ifmmode~\mbox{\c{}}\else \c{}\fi{}ka\l{}a}},\ }\bibfield  {title} {\bibinfo {title} {Temperature and disorder chaos in three-dimensional {Ising} spin glasses},\ }\href {https://doi.org/10.1103/PhysRevLett.98.017201} {\bibfield  {journal} {\bibinfo  {journal} {Phys. Rev. Lett.}\ }\textbf {\bibinfo {volume} {98}},\ \bibinfo {pages} {017201} (\bibinfo {year} {2007})}\BibitemShut {NoStop}%
\bibitem [{\citenamefont {Fernandez}\ \emph {et~al.}(2013)\citenamefont {Fernandez}, \citenamefont {Martin-Mayor}, \citenamefont {Parisi},\ and\ \citenamefont {Seoane}}]{Fernandez_2013}%
  \BibitemOpen
  \bibfield  {author} {\bibinfo {author} {\bibfnamefont {L.~A.}\ \bibnamefont {Fernandez}}, \bibinfo {author} {\bibfnamefont {V.}~\bibnamefont {Martin-Mayor}}, \bibinfo {author} {\bibfnamefont {G.}~\bibnamefont {Parisi}},\ and\ \bibinfo {author} {\bibfnamefont {B.}~\bibnamefont {Seoane}},\ }\bibfield  {title} {\bibinfo {title} {Temperature chaos in 3d {Ising} spin glasses is driven by rare events},\ }\href {https://doi.org/10.1209/0295-5075/103/67003} {\bibfield  {journal} {\bibinfo  {journal} {Europhys. Lett.}\ }\textbf {\bibinfo {volume} {103}},\ \bibinfo {pages} {67003} (\bibinfo {year} {2013})}\BibitemShut {NoStop}%
\bibitem [{\citenamefont {Wang}\ \emph {et~al.}(2015)\citenamefont {Wang}, \citenamefont {Machta},\ and\ \citenamefont {Katzgraber}}]{Wang2015}%
  \BibitemOpen
  \bibfield  {author} {\bibinfo {author} {\bibfnamefont {W.}~\bibnamefont {Wang}}, \bibinfo {author} {\bibfnamefont {J.}~\bibnamefont {Machta}},\ and\ \bibinfo {author} {\bibfnamefont {H.~G.}\ \bibnamefont {Katzgraber}},\ }\bibfield  {title} {\bibinfo {title} {Chaos in spin glasses revealed through thermal boundary conditions},\ }\href {https://doi.org/10.1103/PhysRevB.92.094410} {\bibfield  {journal} {\bibinfo  {journal} {Phys. Rev. B}\ }\textbf {\bibinfo {volume} {92}},\ \bibinfo {pages} {094410} (\bibinfo {year} {2015})}\BibitemShut {NoStop}%
\bibitem [{\citenamefont {Billoire}\ \emph {et~al.}(2018)\citenamefont {Billoire}, \citenamefont {Fernandez}, \citenamefont {Maiorano}, \citenamefont {Marinari}, \citenamefont {Martin-Mayor}, \citenamefont {Moreno-Gordo}, \citenamefont {Parisi}, \citenamefont {Ricci-Tersenghi},\ and\ \citenamefont {Ruiz-Lorenzo}}]{Billoire2018}%
  \BibitemOpen
  \bibfield  {author} {\bibinfo {author} {\bibfnamefont {A.}~\bibnamefont {Billoire}}, \bibinfo {author} {\bibfnamefont {L.~A.}\ \bibnamefont {Fernandez}}, \bibinfo {author} {\bibfnamefont {A.}~\bibnamefont {Maiorano}}, \bibinfo {author} {\bibfnamefont {E.}~\bibnamefont {Marinari}}, \bibinfo {author} {\bibfnamefont {V.}~\bibnamefont {Martin-Mayor}}, \bibinfo {author} {\bibfnamefont {J.}~\bibnamefont {Moreno-Gordo}}, \bibinfo {author} {\bibfnamefont {G.}~\bibnamefont {Parisi}}, \bibinfo {author} {\bibfnamefont {F.}~\bibnamefont {Ricci-Tersenghi}},\ and\ \bibinfo {author} {\bibfnamefont {J.~J.}\ \bibnamefont {Ruiz-Lorenzo}},\ }\bibfield  {title} {\bibinfo {title} {{Dynamic variational study of chaos: Spin glasses in three dimensions}},\ }\href {https://doi.org/10.1088/1742-5468/aaa387} {\bibfield  {journal} {\bibinfo  {journal} {J. Stat. Mech.}\ }\textbf {\bibinfo {volume} {2018}},\ \bibinfo {pages} {033302} (\bibinfo {year} {2018})}\BibitemShut {NoStop}%
\bibitem [{\citenamefont {Baity-Jesi}\ \emph {et~al.}(2021)\citenamefont {Baity-Jesi}, \citenamefont {Calore}, \citenamefont {Cruz}, \citenamefont {Fernandez}, \citenamefont {Gil-Narvion}, \citenamefont {{Gonzalez-Adalid Pemartin}}, \citenamefont {Gordillo-Guerrero}, \citenamefont {I{\~{n}}iguez}, \citenamefont {Maiorano}, \citenamefont {Marinari}, \citenamefont {Martin-Mayor}, \citenamefont {Moreno-Gordo}, \citenamefont {Mu{\~{n}}oz-Sudupe}, \citenamefont {Navarro}, \citenamefont {Paga}, \citenamefont {Parisi}, \citenamefont {Perez-Gaviro}, \citenamefont {Ricci-Tersenghi}, \citenamefont {Ruiz-Lorenzo}, \citenamefont {Schifano}, \citenamefont {Seoane}, \citenamefont {Tarancon}, \citenamefont {Tripiccione},\ and\ \citenamefont {Yllanes}}]{Baity-Jesi2021}%
  \BibitemOpen
  \bibfield  {author} {\bibinfo {author} {\bibfnamefont {M.}~\bibnamefont {Baity-Jesi}}, \bibinfo {author} {\bibfnamefont {E.}~\bibnamefont {Calore}}, \bibinfo {author} {\bibfnamefont {A.}~\bibnamefont {Cruz}}, \bibinfo {author} {\bibfnamefont {L.~A.}\ \bibnamefont {Fernandez}}, \bibinfo {author} {\bibfnamefont {J.~M.}\ \bibnamefont {Gil-Narvion}}, \bibinfo {author} {\bibfnamefont {I.}~\bibnamefont {{Gonzalez-Adalid Pemartin}}}, \bibinfo {author} {\bibfnamefont {A.}~\bibnamefont {Gordillo-Guerrero}}, \bibinfo {author} {\bibfnamefont {D.}~\bibnamefont {I{\~{n}}iguez}}, \bibinfo {author} {\bibfnamefont {A.}~\bibnamefont {Maiorano}}, \bibinfo {author} {\bibfnamefont {E.}~\bibnamefont {Marinari}}, \bibinfo {author} {\bibfnamefont {V.}~\bibnamefont {Martin-Mayor}}, \bibinfo {author} {\bibfnamefont {J.}~\bibnamefont {Moreno-Gordo}}, \bibinfo {author} {\bibfnamefont {A.}~\bibnamefont {Mu{\~{n}}oz-Sudupe}}, \bibinfo {author} {\bibfnamefont {D.}~\bibnamefont {Navarro}}, \bibinfo {author} {\bibfnamefont
  {I.}~\bibnamefont {Paga}}, \bibinfo {author} {\bibfnamefont {G.}~\bibnamefont {Parisi}}, \bibinfo {author} {\bibfnamefont {S.}~\bibnamefont {Perez-Gaviro}}, \bibinfo {author} {\bibfnamefont {F.}~\bibnamefont {Ricci-Tersenghi}}, \bibinfo {author} {\bibfnamefont {J.~J.}\ \bibnamefont {Ruiz-Lorenzo}}, \bibinfo {author} {\bibfnamefont {S.~F.}\ \bibnamefont {Schifano}}, \bibinfo {author} {\bibfnamefont {B.}~\bibnamefont {Seoane}}, \bibinfo {author} {\bibfnamefont {A.}~\bibnamefont {Tarancon}}, \bibinfo {author} {\bibfnamefont {R.}~\bibnamefont {Tripiccione}},\ and\ \bibinfo {author} {\bibfnamefont {D.}~\bibnamefont {Yllanes}},\ }\bibfield  {title} {\bibinfo {title} {{Temperature chaos is present in off-equilibrium spin-glass dynamics}},\ }\href {https://doi.org/10.1038/s42005-021-00565-9} {\bibfield  {journal} {\bibinfo  {journal} {Commun. Phys.}\ }\textbf {\bibinfo {volume} {4}},\ \bibinfo {pages} {74} (\bibinfo {year} {2021})}\BibitemShut {NoStop}%
\bibitem [{\citenamefont {Nishimori}(1993)}]{Nishimori1993}%
  \BibitemOpen
  \bibfield  {author} {\bibinfo {author} {\bibfnamefont {H.}~\bibnamefont {Nishimori}},\ }\bibfield  {title} {\bibinfo {title} {{Optimum decoding temperature for error-correcting codes}},\ }\href {https://doi.org/10.1143/JPSJ.62.2973} {\bibfield  {journal} {\bibinfo  {journal} {J. Phys. Soc. Jpn.}\ }\textbf {\bibinfo {volume} {62}},\ \bibinfo {pages} {2973} (\bibinfo {year} {1993})}\BibitemShut {NoStop}%
\bibitem [{\citenamefont {Bi}\ and\ \citenamefont {Tong}(2015)}]{Bi2015}%
  \BibitemOpen
  \bibfield  {author} {\bibinfo {author} {\bibfnamefont {S.}~\bibnamefont {Bi}}\ and\ \bibinfo {author} {\bibfnamefont {N.-H.}\ \bibnamefont {Tong}},\ }\bibfield  {title} {\bibinfo {title} {Monte carlo algorithm for free energy calculation},\ }\href {https://doi.org/10.1103/PhysRevE.92.013310} {\bibfield  {journal} {\bibinfo  {journal} {Phys. Rev. E}\ }\textbf {\bibinfo {volume} {92}},\ \bibinfo {pages} {013310} (\bibinfo {year} {2015})}\BibitemShut {NoStop}%
\bibitem [{\citenamefont {Yasuda}\ and\ \citenamefont {Takahashi}(2022)}]{Yasuda2022}%
  \BibitemOpen
  \bibfield  {author} {\bibinfo {author} {\bibfnamefont {M.}~\bibnamefont {Yasuda}}\ and\ \bibinfo {author} {\bibfnamefont {C.}~\bibnamefont {Takahashi}},\ }\bibfield  {title} {\bibinfo {title} {Free energy evaluation using marginalized annealed importance sampling},\ }\href {https://doi.org/10.1103/PhysRevE.106.024127} {\bibfield  {journal} {\bibinfo  {journal} {Phys. Rev. E}\ }\textbf {\bibinfo {volume} {106}},\ \bibinfo {pages} {024127} (\bibinfo {year} {2022})}\BibitemShut {NoStop}%
\bibitem [{\citenamefont {Okunishi}\ \emph {et~al.}(2022)\citenamefont {Okunishi}, \citenamefont {Nishino},\ and\ \citenamefont {Ueda}}]{Okunishi2022}%
  \BibitemOpen
  \bibfield  {author} {\bibinfo {author} {\bibfnamefont {K.}~\bibnamefont {Okunishi}}, \bibinfo {author} {\bibfnamefont {T.}~\bibnamefont {Nishino}},\ and\ \bibinfo {author} {\bibfnamefont {H.}~\bibnamefont {Ueda}},\ }\bibfield  {title} {\bibinfo {title} {Developments in the tensor network - from statistical mechanics to quantum entanglement},\ }\href {https://doi.org/10.7566/JPSJ.91.062001} {\bibfield  {journal} {\bibinfo  {journal} {J. Phys. Soc. Jpn.}\ }\textbf {\bibinfo {volume} {91}},\ \bibinfo {pages} {062001} (\bibinfo {year} {2022})}\BibitemShut {NoStop}%
\bibitem [{\citenamefont {Liu}\ \emph {et~al.}(2023)\citenamefont {Liu}, \citenamefont {Ye}, \citenamefont {Yu}, \citenamefont {Duan},\ and\ \citenamefont {Deng}}]{Liu2023}%
  \BibitemOpen
  \bibfield  {author} {\bibinfo {author} {\bibfnamefont {Z.}~\bibnamefont {Liu}}, \bibinfo {author} {\bibfnamefont {Q.}~\bibnamefont {Ye}}, \bibinfo {author} {\bibfnamefont {L.-W.}\ \bibnamefont {Yu}}, \bibinfo {author} {\bibfnamefont {L.~M.}\ \bibnamefont {Duan}},\ and\ \bibinfo {author} {\bibfnamefont {D.-L.}\ \bibnamefont {Deng}},\ }\bibfield  {title} {\bibinfo {title} {{Theory on variational high-dimensional tensor networks}},\ }\href {http://arxiv.org/abs/2303.17452} {\bibfield  {journal} {\bibinfo  {journal} {arXiv:2303.17452}\ } (\bibinfo {year} {2023})}\BibitemShut {NoStop}%
\end{thebibliography}
%

\end{document}